\documentstyle[aps,preprint,eqsecnum]{revtex}
\tighten
\def\frc#1#2{{\textstyle{{#1 \over #2}}}}

\begin{document}
\title{Method to compute the stress-energy tensor for the massless spin~$\frac12$ field in a general static spherically symmetric spacetime}
\author{Peter B. Groves\thanks{Electronic address: {\tt grovpb4@wfu.edu}}, Paul R. Anderson\thanks{Electronic address: {\tt anderson@wfu.edu}}, and Eric D. Carlson\thanks{Electronic address: {\tt ecarlson@wfu.edu}}}
\address{Department of Physics, Wake Forest University, Winston-Salem, North Carolina, 27109}
\maketitle
\begin{abstract}
 A method for computing the stress-energy tensor for the quantized, massless, spin $\frac12$ field in a general static spherically symmetric spacetime is presented. The field can be in a zero temperature state or a non-zero temperature thermal state.  An expression for the full renormalized stress-energy tensor is derived.  It consists of a sum of two tensors both of which are conserved.  One tensor is written in terms of the modes of the quantized field and has zero trace.  
In most cases it must be computed numerically.  The other tensor does not explicitly depend on the modes and has a trace equal to the trace anomaly. It can be used as an analytic approximation for the stress-energy tensor and is equivalent to other approximations that have been made for the stress-energy tensor of the massless spin $\frac12$ field in static spherically symmetric spacetimes.

\end{abstract}

\vspace{.5in}
\section{Introduction}

One of the most useful tools in the study of quantum effects in curved spacetime is the stress-energy tensor $\langle T_{\mu\nu} \rangle$.  The stress-energy tensor gives information about the effects the spacetime has on the quantized fields that propagate in it and also provides a source for that geometry via the semiclassical backreaction equations
\begin{equation}
G_{\mu\nu} = 8 \pi \langle T_{\mu\nu} \rangle \;.
\label{eq:semiclassical}
\end{equation}
Unlike alternative approaches, such as using Bogolubov transformations or particle detectors, the stress-energy
tensor includes the effects of both particle production and vacuum polarization.  It therefore does not
depend on the distinction between these two effects which is often blurred in curved spacetime.

In four-dimensions, the computation of $\langle T_{\mu\nu} \rangle$  
for a quantized field in a given curved spacetime background is at best nontrivial
and at worst almost impossible, given current technology.  There are four types
of approaches that have been used to compute the stress-energy tensor
for a quantized field in a static spherically symmetric spacetime.  One is to
constrain the form of the stress-energy tensor by 
integrating the conservation equation and using the symmetry properties of the 
state that the field is in.  If the field is conformally invariant the fact that
its trace is known can also be used.  This approach has been used in Schwarzschild
spacetime~\cite{cf} for fields in states that are static and spherically symmetric
and for conformally invariant fields in such states in a global monopole spacetime~\cite{hiscock,ml}.

A second approach is to derive an analytic approximation for the stress-energy tensor.
This approach is sometimes combined with the previous one.
Analytic approximations have been derived for conformally invariant fields of spin $0$, $\frc12$, and 1
in Schwarzschild spacetime~\cite{page,hc,howard,bop,vaz1,vaz2,nugayev,mat1,mat2,mat3,mat4,visser,mat5,mat6}, 
in Reissner-Nordstr\"{o}m
and extreme Reissner-Nordstr\"{o}m spacetimes~\cite{zannias,huang}, and in a general static spacetime~\cite{fz1}. 
An approximation has also been derived for massless scalar fields with arbitrary curvature coupling in
a general static spherically symmetric spacetime~\cite{ahs}.
The approximations in Schwarzschild and Reissner-Nordstr\"{o}m spacetimes are mostly for the Hartle-Hawking-Israel~\cite{hh,israel} state 
but in some cases are also for the Unruh~\cite{unruh} and Boulware~\cite{boulware1} states.  The more general approximations are
for the vacuum state and a thermal state at an arbitrary temperature.
For massive fields the DeWitt-Schwinger
expansion~\cite{dewitt} can often be used to obtain a good approximation for the stress-energy tensor.  This approximation
has been derived for massive scalar fields in Schwarzschild spacetime~\cite{frolov,fz2,fz3,fz4}, in a general
static spherically symmetric spacetime~\cite{ahs,mat6}, and in a general spacetime~\cite{mat7}.  
It has also been derived for massive spin $\frc12$ and $1$ fields 
in Reissner-Nordstr\"{o}m spacetimes~\cite{mat8}.  The
DeWitt-Schwinger approximation is independent of the state of the field.

A third approach is to numerically compute the stress-energy tensor.  This has been done
for conformally invariant spin $0$ and spin $1$ fields in 
Schwarzschild spacetime on 
the event horizon~\cite{candelas} and outside of the event horizon~\cite{fawcett,hc,jo,ahs,jmo}.
Numerical calculations of the stress-energy tensor for both massive and massless scalar fields with arbitrary
curvature couplings 
in Schwarzschild and Reissner-Nordstr\"{o}m spacetimes outside the event horizon have also been done~\cite{ahs,ahl}. 
Finally a numerical computation of the stress-energy tensor for the massless spin $\frc12$ field has
been done in a global monopole spacetime~\cite{bbk}. 

The fourth approach is the most accurate and the most difficult.  It is to analytically compute
the full renormalized stress-energy tensor.  This has been done for the
special case of conformally invariant fields in conformally flat spacetimes~\cite{bc,bunch} and for scalar fields with arbitrary mass and curvature coupling in de Sitter 
space~\cite{d-c,bu-da,a-f,f,k-g}.  The stress-energy tensor for the massless spin $1$ field has
also been computed analytically on the event horizon of a Schwarzschild black hole\cite{elster}.  To our knowledge 
no other exact analytic calculations
of the stress-energy tensor in static spherically symmetric spacetimes have been published.

Clearly only a little has been done regarding the computation of the full renormalized stress-energy
tensor for the massless spin $\frc12$ field.  In particular
the only calculation of the full renormalized stress-energy tensor in a black hole spacetime is on the event horizon of the extreme Reissner-Nordstr\"{o}m black hole~\cite{ahl}.  This was possible because the geometry near the
horizon asymptotically approaches the conformally flat Bertotti-Robinson geometry~\cite{bertotti,robinson,carter}.  

In this paper we present a practical method that can be used to numerically compute the stress-energy tensor for the massless spin $\frc12$ field in a general static spherically symmetric spacetime.  
The method makes use of the Euclidean Green's function for this field.  Our approach closely follows that of Anderson, Hiscock, and Samuel \cite{ahs} which in turn was built on that of Howard and Candelas \cite{hc,howard}. As in those papers, the Euclidean Green's function is constructed using a method similar to that of Candelas~\cite{candelas}, and point splitting is used to renormalize the stress-energy tensor.  The point splitting counterterms for spin $0$, $\frc12$, and $1$ fields have been computed by Christensen~\cite{christensen1,christensen2}.

Our main result is the derivation of a renormalized expression for the stress-energy tensor of the 
massless spin $\frc12$ field
in a general static spherically symmetric spacetime.  The field can be either in a vacuum state or in 
a thermal state at an arbitrary temperature.  One advantage of our expression is that one does not need to work
with a vierbein or any Dirac matrices to use it.  These were involved in the derivation of the expression but no longer
are a part of it.  In our derivation the stress-energy tensor is divided into two parts.  One part depends on sums and
integrals over radial mode functions.  This part usually needs to be computed numerically.  The other part consists of an analytic tensor with a trace equal to the trace anomaly for the spin $\frc12$ field.  Since both parts are separately conserved, the analytic part can be used as an analytic approximation for the stress-energy tensor.  It is equivalent to the approximation of Brown, Ottewill, and Page \cite{bop} in Schwarzschild spacetime, and to that of Frolov and Zel'nikov \cite{fz1} in a general static spherically symmetric spacetime for certain values of the three arbitrary coefficients in their expressions.

In Section \ref{rev_spinors}, we review the formalism used to describe spinors in curved space and its application to static spherically symmetric spacetimes.  In Section \ref{euc_green's_func} we construct an expression for the Euclidean Green's function for the massless spin $\frc12$ field in a general static spherically symmetric spacetime.  
The WKB approximation for the radial mode functions is introduced in Section \ref{WKB} and 
the point splitting renormalization procedure is reviewed in Section \ref{rev_pt_split}.  In Section \ref{Tmn_unren} an expression for the unrenormalized stress-energy tensor is derived.   The renormalized stress-energy tensor is derived in Section \ref{Tmnren}, and a part that can be used as an analytic approximation is displayed.  Our results are summarized and discussed in Section \ref{Discussion}.  Various technical results are derived in the Appendices.

Our conventions are those of Misner, Thorne, and Wheeler \cite{mtw} and we use units such that $\hbar = c = G = k_B = 1$.  Greek indices are used to refer to general spacetime coordinates and Latin indices refer to locally inertial coordinates.

\section{Review of Spinors in Curved Spacetime}
\label{rev_spinors}

The study of spin $\frc12$ fields in curved spacetime appears to have begun with Fock and Ivanenko~\cite{fi} who derived
the Dirac equation in a general spacetime.  Brill and Wheeler \cite{brill-wheeler} adapted the formalism of spinors in curved space to static spherically symmetric spacetimes and derived the
specific form of the Dirac equation in these spacetimes.  Green's functions for the spin $\frc12$ field have been
derived by Boulware \cite{boulware2} for Schwarzschild spacetime and by Bezerra de Mello, Bezerra, and Khusnutdinov~\cite{bbk} in a global monopole spacetime.

In this section we review the formalism most commonly used to treat spin $ \frc12 $ fields in a general curved spacetime.
We then derive an explicit form for the Dirac equation in Euclidean space and discuss its solutions.

\subsection{General Results}
\label{gen_results}

The spin of a field $\psi$ is defined by its behavior under a Lorentz transformation.  
An infinitesimal Lorentz transformation in flat space can be written in the form 
\begin{equation}
{\Lambda^a}_b = {\delta^a}_b + {\omega^a}_b
\end{equation}
with $ {\omega^a}_b $ an arbitrary infinitesimal antisymmetric tensor.  The transformation law for a field $\psi$ of arbitrary spin is \cite{weinberg} 
\begin{eqnarray}
\label{eq:Lor_trans}
	\psi \rightarrow D ( \Lambda ) \psi = \exp\left(\frc12 \omega_{ab} \sigma^{ab}\right) \psi
\end{eqnarray}
where $D$ denotes a representation of the Lorentz group and the matrices $\sigma^{ab}$ are the 
generators of this representation. For an infinitesimal Lorentz transformation,
\begin{equation}
	D ( \Lambda ) = 1+ \frc12 \omega_{ab} \sigma^{ab}  \; .
\end{equation}
The matrices $\sigma^{ab}$ satisfy the commutation relations
\begin{eqnarray}
\label{eq:sigma_condition}
	\left[ \sigma^{ab} , \sigma^{cd} \right] = \eta^{cb} \sigma^{ad} - \eta^{ca} \sigma^{bd} + \eta^{db} \sigma^{ca} - \eta^{da} \sigma^{cb} \; .
\end{eqnarray}
Here $\eta_{ab}$ is the usual Minkowski metric in Cartesian coordinates.  For spin $\frc12$ fields~\cite{birrell-davies}
\begin{equation}
\label{eq:sigma_ab_lor}
\sigma^{ab} = - \frc14\left[\gamma^a,\gamma^b \right] \;,
\end{equation}
where the $\gamma^a$ are $4\times4$ matrices satisfying the anticommutation relations\footnote{Had we chosen $ \{\gamma^a,\gamma^b\} = 2\eta^{ab}{\bf 1} $, we would have chosen $ \sigma^{ab} =  \frc14\left[\gamma^a,\gamma^b \right] $ .}

\begin{equation}
\{\gamma^a,\gamma^b\} = -2\eta^{ab}{\bf 1} \;.
\end{equation}
For this work, we use the Dirac representation \cite{itzykson-zuber}:
\begin{eqnarray}
	\gamma^0 = \pmatrix{{\bf 1} & {\bf 0} \cr {\bf 0} & -{\bf 1} },
	\hspace{2cm} \gamma^i = \pmatrix{ {\bf 0} & \sigma_i \cr  -\sigma_i & {\bf 0 }}
	\hspace{2cm} i = 1,2,3 \;.
\end{eqnarray}
Here the $ \sigma_i $ are the usual Pauli matrices
\begin{equation}
\label{eq:pauli_matrices}
              \sigma_1= \pmatrix{ 0 & 1 \cr 1 & 0} \; ,
\hspace{.5cm} \sigma_2= \pmatrix{0 & -i \cr i & 0} \; ,
\hspace{.5cm} \sigma_3= \pmatrix{1 & 0 \cr 0 & -1} \; .
\end{equation}

In curved spacetime, it is useful to use the vierbein or tetrad formalism \cite{weinberg}.  In this formalism, one sets up a locally inertial frame at every point.  The coordinate transformation between the general spacetime coordinates and the 
coordinates for the frame that is locally inertial at a given point $x$ is
given by the vierbein functions $ {e_a}^\mu(x) $.  As a result the relationship between the locally inertial
metric $\eta_{ab}$ and the spacetime metric $g_{\mu\nu}$ at $x$ is
\begin{equation}
 \eta_{ab} = g_{\mu \nu}(x) {e_a}^\mu(x) {e_b}^{\nu}(x) \;.
\label{eq:vierbein2}  
\end{equation} 
Note that the locally inertial coordinate indices are raised and lowered with $ \eta_{ab} $ and the general coordinate indices with $ g_{\mu \nu} $.  It is easy to show that the vierbein functions also satisfy
\begin{eqnarray}
{e_a}^\mu {e^b}_\mu &=& \delta_a^b  \nonumber \\
{e_a}^\mu {e^a}_{\nu} &=& \delta^\mu_{\nu} \; .
\label{eq:vierbein1}
\end{eqnarray}

Using the vierbein formalism, one is able to maintain Lorentz covariance and general coordinate covariance through the definition of an appropriate covariant derivative \cite{weinberg}.  One accomplishes this by requiring that the spinor covariant derivatives behave in the same manner in curved space as in flat space.  In flat space the gradient of a spinor is both a spinor and a vector.  This implies that under a Lorentz transformation $ \partial_{\alpha} \psi \rightarrow {\Lambda_{\alpha}}^{\beta} D \left( \Lambda \right) \partial_{\beta} \psi $.  In curved space, the covariant derivative is defined in such a way that this relationship is preserved~\cite{weinberg,birrell-davies}.  Thus
\begin{eqnarray}
\label{eq:covariant_der_def}
	\nabla_{a} \psi(x) \rightarrow {\Lambda_a}^b \left( x \right) D \left( \Lambda \left( x \right) \right) \nabla_{b} \psi(x) 
\end{eqnarray}
where $\nabla_{a} = {e_a}^\mu \nabla_\mu$ and
 $ \Lambda $ is a function of $ x $, since the locally inertial frame is different, in general, at different points in the spacetime.
Equation (\ref{eq:covariant_der_def}) is satisfied if we define
\begin{equation}
\label{eq:der_def}
     \nabla_\mu \equiv \partial_\mu + \Gamma_\mu 
\end{equation}
with
\begin{eqnarray}
\label{eq:Gamma_def}
	\Gamma_\mu \equiv \frc12 \sigma^{ab} {e_a}^\nu e_{b\nu; \mu} \; .
\end{eqnarray}
Here and throughout the semicolon denotes the covariant derivative of a vector or tensor.  It acts on the spacetime indices (denoted by Greek letters).

With the vierbein formalism, the Dirac equation for a massless spin $\frc12$ field in an arbitrary spacetime
can be written as \cite{birrell-davies}
\begin{equation}
\label{eq:dirac_equation}
     i \gamma^\mu \nabla_\mu \psi = 0 \; , 
\end{equation}
where $\gamma^\mu = \gamma^a{e_a}^\mu$ .

To calculate $ \langle T_{\mu \nu} \rangle $ in a static spacetime, it is advantageous to work in Euclidean space instead of Lorentzian space.  To compute covariant derivatives in Euclidean space, one may repeat the above steps, but this time preserving covariance under a Euclidean rotation instead of a Lorentz transformation.  An infinitesimal rotation ${\cal R}_{ab} $ in Euclidean space can be written as
\begin{equation}
	{\cal{R}}_{ab} = \delta_{ab} + \omega^E_{ab}
\end{equation}
where $\omega^E_{ab}$ is an arbitrary antisymmetric tensor.  Under a rotation $ {\cal{R}} $, a field $ \psi $ of arbitrary spin has the transformation
\begin{equation}
\label{eq:D_rotation}
	\psi \rightarrow D ( {\cal{R}} ) \psi = \exp\left(\frc12 \omega^E_{ab}\sigma_E^{ab}\right) \psi \; .
\end{equation}
For an infinitesimal rotation this becomes
\begin{equation}
\label{eq:infinitesimal_rotation}
	D ( {\cal{R}} ) = 1+ \frc12 \omega^E_{ab} \sigma^{ab}_E  \; .
\end{equation}
The $\sigma^{ab}_E$ must be generators of a representation of the Euclidean group (in four dimensions).  They satisfy
the commutation relations
\begin{eqnarray}
\label{eq:sigmaE_condition}
	\left[ \sigma^{ab}_E , \sigma^{cd}_E \right] = \delta^{cb} \sigma^{ad}_E - \delta^{ca} \sigma^{bd}_E + \delta^{db} \sigma^{ca}_E - \delta^{da} \sigma^{cb}_E \; .
\end{eqnarray}
For spin $\frc12$ fields 
\begin{eqnarray}
\label{eq:sigma_ab_euc}
	\sigma^{ab}_E & = & \frc14\left[\gamma^a_E,\gamma^b_E \right]
\end{eqnarray}
with
\begin{equation}
	\{\gamma_E^a,\gamma_E^b\} = 2 \delta^{ab} \;.
\end{equation}
It is easy to show that this anticommutation relation is satisfied if we make the identification
\begin{eqnarray}
\label{eq:gamma_euclidean}
	\gamma_E^0 \equiv \gamma^0 \hspace{2cm} \gamma_E^i \equiv -i \gamma^i \; .
\end{eqnarray}

Note that the relationship between the vierbein and the metric in Eq.\ (\ref{eq:vierbein2}) allows us to use the same vierbein for both the Euclidean and Lorentzian sectors of a given static spacetime.
The Euclidean covariant derivative can then be defined analogously to the Lorentzian one, by requiring that the covariant derivative of a spinor in curved space behave in the same manner as in flat space:
\begin{eqnarray}
\label{eq:covariant_der_def_E}
	\nabla^E_{\mu} \psi \left( x \right) \rightarrow {\cal{R}_\alpha}^\beta \left( x \right) D \left( {\cal{R}} \left( x \right) \right) \nabla^E_\beta \psi \left( x \right) \; .
\end{eqnarray}
Eq.\ (\ref{eq:covariant_der_def_E}) is satisfied if 
\begin{equation}
\label{eq:nabla_E_definition}
     \nabla^E_\mu \equiv\partial^E_\mu + \Gamma^E_\mu
\end{equation} 
with 
\begin{eqnarray}
\label{eq:Gamma_mu_E}
	\Gamma^E_\mu \equiv \frc12 \sigma^{ab}_E {e_a}^\nu e_{b\nu; \mu} \; .
\end{eqnarray}

Then the Dirac equation for a massless spin $\frc12$ field in an arbitrary Euclidean spacetime is
\begin{equation}
\label{eq:dirac_equation_euc}
     - \gamma^\mu_E \nabla^E_\mu \psi = 0 \;.
\end{equation}

\subsection{Static Spherically Symmetric Spacetimes}
\label{ssss}

For a general static spherically symmetric spacetime the metric can be written in the form
\begin{equation}
     ds^2=-f(r) \, dt^2 + h(r) \, dr^2 + r^2 \, d \theta^2 +r^2 \sin^2(\theta) \, d\phi^2
\label{eq:ssss_metric}
\end{equation}
The transformation to Euclidean space is made by setting $t = -i \tau$ in Eq.\ (\ref{eq:ssss_metric}).
One vierbein that is appropriate for this metric is 
\begin{eqnarray}
\label{eq:vierbein}
     {e_a}^\mu=
     \pmatrix{f^{-1/2} & 0 & 0 & 0 \cr
  0 & h^{-1/2} \sin\theta \cos\phi  & r^{-1}\cos\theta \cos\phi  &  -r^{-1}\csc\theta\sin\phi \cr
  0 & h^{-1/2} \sin\theta \sin\phi  & r^{-1}\cos\theta \sin\phi  &  r^{-1}\csc\theta \cos\phi \cr
  0 & h^{-1/2} \cos\theta   & -r^{-1} \sin\theta & 0}  \; .
\end{eqnarray}
The rows of this matrix are specified by the index $a$ and the columns by the index $\mu$.
One may derive explicit forms for the spin connection matrices by substituting Eqs.\ (\ref{eq:sigma_ab_lor}) and (\ref{eq:vierbein}) into Eq.\ (\ref{eq:Gamma_def}).  
The spin connections for Euclidean space can be found by substituting Eqs.\ (\ref{eq:sigma_ab_euc}), (\ref{eq:gamma_euclidean}), and (\ref{eq:vierbein}) into Eq.\ (\ref{eq:Gamma_mu_E}).  In both cases the results are
\begin{eqnarray}
\label{eq:Gammas}
\Gamma_t &=& i \Gamma_{\tau} = {f' \gamma^0 \tilde{\gamma}\over 4 f^{1/2} h^{1/2}} \nonumber \\
\Gamma_r &=& 0 \nonumber \\
\Gamma_\theta &=& \frc12(1-h^{-1/2})\left(\gamma^1 \gamma^3\cos\phi + \gamma^2\gamma^3\sin\phi\right) \nonumber \\
\Gamma_\phi &=& \frc12(h^{-1/2}-1)\left(\gamma^1 \gamma^2\sin\theta + \gamma^1\gamma^3 \cos\theta \, \sin\phi -\gamma^2\gamma^3 \cos\theta\, \cos\phi\right)\sin\theta 
\end{eqnarray}
with
\begin{equation}
\tilde{\gamma} \equiv \sin{\theta} \cos{\phi} \, \gamma^1 + \sin{\theta} \sin{\phi} \, \gamma^2
     + \cos{\theta} \, \gamma^3 \;.  
\end{equation}
Substituting Eqs.\ (\ref{eq:Gammas}) into Eq.\ (\ref{eq:dirac_equation_euc}) results in the following form
for the Euclidean space Dirac equation:
\begin{eqnarray}
	- {\gamma^0 \over f^{1/2}}\partial_\tau\psi + { i\, \tilde{\gamma} \over r f^{1/4}h^{1/2}} \partial_r \left(r f^{1/4}\psi\right) - {i \,  \tilde{\gamma} \over r} \left(\vec \Sigma \cdot \vec L + {\bf 1}\right) \psi =0 \; ,
\label{eq:dirac_euclidean}
\end{eqnarray} 
where in the Dirac representation,
\begin{eqnarray}
\vec \Sigma = \pmatrix{ \vec \sigma & {\bf 0} \cr {\bf 0} & \vec \sigma } \; ,
\end{eqnarray}
and $\vec L$ is the standard angular momentum operator with components
\begin{eqnarray}
\label{eq:ang_op}
\nonumber L_1 &=& i \sin\phi \, \partial_\theta + i \cot\theta \, \cos \phi \, \partial_\phi \\
\nonumber L_2 &=& - i \cos\phi \, \partial_\theta + i \cot\theta \, \sin \phi \, \partial_\phi \\
          L_3 &=& - i \partial_\phi \; .
\end{eqnarray}

Using separation of variables, one finds that, as in the Lorentzian case~\cite{brill-wheeler,thaller,bjorken},
 there are two types of solutions to the Euclidean Dirac equation.  When this equation is written in
the form (\ref{eq:dirac_euclidean}), they are
 \begin{eqnarray}
\label{eq:psi2}
    \nonumber \psi_1(x)&=& {e^{i \omega \tau} \over r f^{1/4}} \pmatrix{F_{\omega,j}(r)\Psi^m_{j,+}(\theta,\phi) \vspace{.2cm} \cr G_{\omega,j}(r)\Psi^m_{j,-}(\theta,\phi) } \label{eq:psi1} \; , \\ [.3cm] 
    \psi_2(x)&=& {e^{i \omega \tau} \over r f^{1/4}} \pmatrix{G_{\omega,j}(r)\Psi^m_{j,-}(\theta,\phi) \vspace{.2cm} \cr F_{\omega,j}(r)\Psi^m_{j,+}(\theta,\phi) } \; .
\end{eqnarray}
The angular functions $ \Psi_{j, \, \pm}^m $ are eigenfunctions of the operator ${\bf k} = \vec \sigma \cdot \vec L + {\bf 1} $, and satisfy \cite{thaller,bjorken}
\begin{eqnarray}
\label{eq:ang_eigen_eqn}
	{\bf k} \Psi_{j,\pm}^m = \pm \left( j+\frc12 \right) \Psi_{j, \pm}^m  \qquad \hbox{and} \qquad \tilde{\sigma} \Psi_{j,\pm}^m = \Psi_{j,\mp}^m ,
\end{eqnarray}
where 
\begin{equation}
\tilde{\sigma} \equiv  \sin \theta \cos \phi \, \sigma_1 + \sin \theta \sin \phi \, \sigma_2 + \cos \theta \, \sigma_3  \; .
\end{equation}
Their explicit form is
\begin{equation}
\Psi_{j,+}^m = \pmatrix{\sqrt{j+m \over 2j}Y_{j{-}\frc12}^{m-\frc12} \vspace{.2cm} \cr \sqrt{j-m \over 2j} Y_{j{-}\frc12}^{m+\frc12}} \qquad \hbox{and} \qquad 
\Psi_{j,-}^m = \pmatrix{\sqrt{j+1-m \over 2j+2}Y_{j{+}\frc12}^{m-\frc12} \vspace{.2cm} \cr - \sqrt{j+1+m \over 2j+2} Y_{j{+}\frc12}^{m+\frc12}} \; 
\label{eq:psij's}
\end{equation}
with $j = \frc12, \frc32, ...$ and $m = -j, -j+1, ..., j$.  
The $Y^{m\pm 1/2}_{j\pm 1/2}$ are the standard spherical harmonics.

Substituting either of $ \psi_1(x) $ or $ \psi_2(x) $ in Eq.\ (\ref{eq:psi1}) into Eq.\ (\ref{eq:dirac_euclidean}) 
yields the following coupled set of mode equations for $F_{\omega\!,j}$  and $G_{\omega\!,j}$:
\begin{mathletters}
\begin{eqnarray}
\label{eq:Feq}
{\omega \over f^{1/2}} F_{\omega,j} &=& {1\over h^{1/2}} \partial_r G_{\omega,j} + {j{+}\frac12 \over r} G_{\omega,j} \; ,\\
\label{eq:Geq}
{\omega \over f^{1/2}} G_{\omega,j} &=& {1\over h^{1/2}} \partial_r F_{\omega,j} - {j{+}\frac12 \over r} F_{\omega,j} \; .
\end{eqnarray}
\end{mathletters}These may be combined to form the second order equations
\begin{mathletters}
\begin{eqnarray}
\label{eq:second_F}
{f^{1/2} \over h^{1/2}} \partial_r \left[ {f^{1/2} \over h^{1/2}} \partial_r F_{\omega,j} \right]
     - F_{\omega,j} {f^{1/2} \over h^{1/2}} \partial_r \left[ \left( j{+}\frc12 \right) {f^{1/2} \over r} \right] 
     - \left( j{+}\frc12 \right)^2 \frac{f}{r^2} F_{\omega,j} - \omega^2 F_{\omega,j} &=& 0 \; ,\\ [.3cm]  
\label{eq:second_G}
{f^{1/2} \over h^{1/2}} \partial_r \left[ {f^{1/2} \over h^{1/2}} \partial_r G_{\omega,j} \right]
     + G_{\omega,j} {f^{1/2} \over h^{1/2}} \partial_r \left[ \left( j{+}\frc12 \right) {f^{1/2} \over r} \right]  
     - \left( j{+}\frc12 \right)^2 \frac{f}{r^2} G_{\omega,j} - \omega^2 G_{\omega,j} &=& 0 \; .
\end{eqnarray}
\end{mathletters}Since $F_{\omega,j}$ and $G_{\omega,j}$ obey a coupled set of linear first order equations, there will
be two linearly independent solutions.  Boundary conditions for the radial modes depend on the physical situation and are usually specified at both large and small values of $ r $.  For example, in a black hole spacetime the modes must be finite at the event horizon and at $r=\infty$.  We denote the solutions that satisfy the boundary condition at small $ r $ by the superscript $ p $.  These are $ F^p_{\omega,j} $ and $ G^p_{\omega,j} $.  Those that satisfy the boundary condition at large  $ r $ are denoted by the superscript $ q $.  These are $ F^q_{\omega,j} $ and $ G^q_{\omega,j} $.  The normalization of these solutions is, for the moment, left unspecified.

\section{The Euclidean Green's Function} 
\label{euc_green's_func}
The Euclidean Green's function for a massless spin $ \frc12 $ field is required 
to be a solution of the equation
\begin{eqnarray}
\label{eq:SE_equation}
     - \gamma^\mu_E \nabla^E_\mu S_E(x,x')= - {\bf 1}{\delta^4(x-x') \over \sqrt{g_E}}
\end{eqnarray}
It must also satisfy the appropriate boundary conditions.    

Using a method similar to that which Candelas \cite{candelas} used to construct the Euclidean Green's function for a scalar
field, we find that the Euclidean Green's function for the spin $\frc12$ field in a zero temperature vacuum state in a static spherically symmetric spacetime can be
written in the form
\begin{eqnarray}
\label{eq:SE}
&& \nonumber S_E(\tau,\vec{x};\tau',\vec x \, ') = i \int_{-\infty}^\infty {d\omega \, \omega e^{i\omega (\tau{-}\tau')}  
     \over 2 \pi rr'f(r)^{1/4}f(r')^{1/4}} \sum_{j=\frac12}^{\infty} \sum_{m=-j}^{j} \\
\nonumber
&& \Biggl\{ \theta(r{-}r') \left[ \pmatrix{ F_{\omega,j}^q(r) \Psi_{j,+}^m(\theta,\phi)  \vspace{.2cm} \cr
                                            G_{\omega,j}^q(r) \Psi_{j,-}^m(\theta,\phi)} 
\otimes \biggl( F_{\omega,j}^p(r') \Psi_{j,+}^{m \dagger}(\theta',\phi') \qquad
                G_{\omega,j}^p(r') \Psi_{j,-}^{m \dagger}(\theta',\phi') \biggr)  \right. \\[.2cm]
\nonumber & & \left.  \hspace{1.5cm} - \pmatrix{ G_{\omega,j}^q(r) \Psi_{j,-}^m(\theta,\phi) \vspace{.2cm} \cr
                                                 F_{\omega,j}^q(r) \Psi_{j,+}^m(\theta,\phi)} 
\otimes \biggl( G_{\omega,j}^p(r') \Psi_{j,-}^{m \dagger}(\theta',\phi') \qquad
                F_{\omega,j}^p(r') \Psi_{j,+}^{m \dagger}(\theta',\phi') \biggr) \right] \\[.2cm]
& & + \, \theta(r'{-}r) \Biggl[ p \leftrightarrow q \Biggr] \Biggr\} \;.
\end{eqnarray}
The notation $\otimes$  denotes a direct product.  The notation
 $ p \leftrightarrow q $ means that the 
superscripts $ p $ and $ q $ on the radial mode functions
$ F_{\omega,j} $ and $ G_{\omega,j} $ should be switched, but everything else, including the arguments of the mode functions, should be left the same as in the previous expression.  Thus the explicit form of the second term in Eq.\ (\ref{eq:SE}) is
\begin{eqnarray}
\nonumber && \theta(r'{-}r) \left[\pmatrix{ F_{\omega,j}^p(r) \Psi_{j,+}^m(\theta,\phi)  \vspace{.2cm} \cr
                                            G_{\omega,j}^p(r) \Psi_{j,-}^m(\theta,\phi)} 
\otimes \biggl( F_{\omega,j}^q(r') \Psi_{j,+}^{m \dagger}(\theta',\phi') \qquad
                G_{\omega,j}^q(r') \Psi_{j,-}^{m \dagger}(\theta',\phi') \biggr)  \right. \\[.2cm]
\nonumber & & \left.  \hspace{1.5cm} - \pmatrix{ G_{\omega,j}^p(r) \Psi_{j,-}^m(\theta,\phi) \vspace{.2cm} \cr
                                                 F_{\omega,j}^p(r) \Psi_{j,+}^m(\theta,\phi)} 
\otimes \biggl( G_{\omega,j}^q(r') \Psi_{j,-}^{m \dagger}(\theta',\phi') \qquad
                F_{\omega,j}^q(r') \Psi_{j,+}^{m \dagger}(\theta',\phi') \biggr) \right] 
\end{eqnarray}
For a thermal state at temperature $T$ the integral in Eq.\ (\ref{eq:SE}) is changed to a sum using the
following rule:
\begin{equation}
 \int_{-\infty}^{\infty} d \omega V(\omega) \rightarrow \kappa \sum_{n=1/2}^\infty \left[V(n \kappa) + V(-n \kappa) \right]
\end{equation}
where $\kappa=2\pi T$ and the sum over $n$ runs over the half integers $\frc12, \frc32, \dots \; $.  This ensures that the spin $\frc12$ Euclidean Green's function is antiperiodic in Euclidean time $\tau$ as it must be for an 
anticommuting field \cite{birrell-davies}.  

An examination of Eq.\ (\ref{eq:SE}) shows that it satisfies the same boundary conditions that the radial modes do at large and small values of $r$.  To verify that Eq.\ (\ref{eq:SE}) can satisfy Eq.\ (\ref{eq:SE_equation}), first note that because Eq.~(\ref{eq:psi2}) is a solution of the homogenous Dirac equation, most of the terms in $\gamma_E^\mu \nabla_\mu^E S_E$ will vanish.  The only non-vanishing terms will be those where the derivative acts on the radial step-functions, yielding factors of $\delta(r-r')$.  Combining terms, and writing out the matrices explicitly, many of the terms will vanish.  If the condition
\begin{equation}
\label{eq:wronskian}
	\omega \; \Bigl[ \, G_{\omega,j}^q (r) F_{\omega,j}^p (r) - F_{\omega,j}^q (r) G_{\omega,j}^p (r) \, \Bigr] = 1
\end{equation}
is imposed, then the integral over $\omega$ can be computed to yield a factor of $\delta(\tau{-}\tau')$.  However, Eq.~(\ref{eq:wronskian}) is just a Wronskian condition for the solutions to Eqs.~(\ref{eq:Feq}) and 
(\ref{eq:Geq}), so the normalization can be chosen so that it is satisfied for all values of $r$.  All that is left is the angular mode sums, which can be computed using the identity
\begin{equation}
\sum_{\ell=0}^\infty \sum_{m=-\ell}^\ell Y_\ell^m(\theta,\phi) Y_\ell^{m\ast}(\theta',\phi')
     = {\delta(\theta-\theta') \; \delta(\phi - \phi') \over \sin\theta}  \; ,
\end{equation}
and the proof that Eq.~(\ref{eq:SE}) satisfies Eq.~(\ref{eq:SE_equation}) is complete.

In spite of the fact that Eq.\ (\ref{eq:SE}) is a solution to (\ref{eq:SE_equation}) with the correct boundary 
conditions, it might appear that our ansatz is not sensible.  After all, in Lorentzian space, we expect expressions like $\psi(x)\overline\psi(x')$, and not $\psi(x)\psi^\dagger(x')$ to appear in the Green's function.  For example, one knows that $\overline \psi \psi $ is invariant under Lorentz transformations.  Since the trace of our $ S_E $ when the points are brought together contains $ \psi^\dagger \psi $ instead of $\overline \psi \psi = \psi^\dagger \gamma^0 \psi$, at first glance it may appear unreasonable.

The reason for the absence of the $\gamma^0$ is that we are working in Euclidean, not Lorentz\-ian space.  A finite rotation in Euclidean space of a field $ \psi $ will transform according to Eq.\ (\ref{eq:D_rotation}).
The $\sigma^{ab}_E$'s in Eq.\ (\ref{eq:D_rotation}) are all anti-Hermitian, as can be seen easily from their definition in Eq.\ (\ref{eq:sigma_ab_euc}).  It follows that
\begin{eqnarray}
\psi^\dagger \rightarrow  \psi^\dagger \exp\left(\frc12 \omega^E_{ab}\sigma^{ab\dagger}_E\right) = \psi^\dagger \exp\left(-\frc12 \omega^E_{ab}\sigma^{ab}_E\right)\; ,
\end{eqnarray}
so that under a local Euclidean rotation, $\psi^\dagger\psi \rightarrow \psi^\dagger \psi$.

In contrast, in Lorentzian space, the $\sigma^{ab}$'s are not all anti-Hermitian.  Specifically,
\begin{equation}
\sigma^{0i\dagger} = \sigma^{0i} \qquad \hbox{while} \qquad \sigma^{ij\dagger} = - \sigma^{ij} \;.
\end{equation}
Here $i$ and $j$ are space indices.  It is easy to demonstrate that $\sigma^{ab\dagger}\gamma^0 = - \gamma^0 \sigma^{ab}$, so that under a Lorentz transformation,
\begin{equation}
\psi^\dagger \gamma^0 \rightarrow \psi^\dagger \exp\left(\frc12\omega_{ab}\sigma^{ab\dagger}\right)\gamma^0 = \psi^\dagger \gamma^0 \exp\left(-\frc12\omega_{ab}\sigma^{ab}\right)
\end{equation}
and thus $\psi^\dagger \gamma^0 \psi \rightarrow \psi^\dagger \gamma^0 \psi$.  Therefore the overall form of our $ S_E $ as the direct product $ \psi(x) \otimes \psi^{\dagger}(x') $ is sensible.  In addition, we show in Appendix \ref{flat_green} that our Green's function is equal to the standard one in flat space.

It is possible to compute the sum over $m$ in Eq.\ (\ref{eq:SE}) by using the addition theorem for spherical harmonics \cite{arfken}:
\begin{eqnarray}
\label{eq:addition_theorum}
	P_{\ell}(\cos\gamma) = \frac{4 \pi}{2 \ell + 1} \sum_{m=-\ell}^{\ell} Y_l^m(\theta, \phi) 
        Y_l^{m \ast}(\theta',\phi') \;,
\end{eqnarray}
with
\begin{eqnarray}
\label{eq:cos_gamma}
	\cos\gamma \equiv \cos\theta \, \cos\theta' + \sin\theta \, \sin\theta' \, \cos(\phi{-}\phi') \; .
\end{eqnarray}
The result is 
\begin{eqnarray}
\label{eq:SE_1}
\nonumber S_E  ( \tau,\vec{x};\tau',\vec x \, ') &=& \int \frac{d \mu\, \omega}{ rr' (f(r)f(r'))^{1/4}}  \times \\
\nonumber &\times& \sum_{\ell=0}^{\infty} \biggl[ \theta(r{-}r') \Bigl\{
     - \sin[\omega(\tau{-}\tau')] \; \gamma^0 \, A_{\omega,\ell}^{q,p}
     + i \cos[\omega(\tau{-}\tau')] \; \tilde\gamma \, B_{\omega,\ell}^{q,p} \Bigr\} \\
& & \qquad + \; \theta(r'{-}r)  \Bigl\{p \leftrightarrow q\Bigr\} \biggr] P_{\ell}(\cos\gamma) 
\end{eqnarray}  
where $\int d\mu$ is defined such that for an arbitrary function $V(\omega)$
\begin{eqnarray}
\nonumber \int d \mu \; V(\omega) &\equiv& {1\over 4\pi^2} \int_0^\infty d\omega V(\omega) \qquad \hbox{if $T=0$} \\ 
&\equiv& \frac{\kappa}{4 \pi^2} \sum_{n=1/2}^\infty V(n \kappa)\qquad \hspace{.1cm}\hbox{if $T>0$} \;,
\end{eqnarray} 
and
\begin{eqnarray}
\label{eq:Aqp}
\nonumber A_{\omega,\ell}^{q,p}&\equiv& F^q_{\omega, \ell{+}\frac12}(r) F^p_{\omega, \ell{+}\frac12}(r') 
          \left[(\ell{+}1){\bf 1} + \vec \Sigma \cdot \vec L\right]
- G^q_{\omega, \ell{-}\frac12}(r) G^p_{\omega, \ell{-}\frac12}(r') \left[\ell{\bf 1} - \vec \Sigma \cdot \vec L\right] \\[.1cm]
B_{\omega,\ell}^{q,p} &\equiv& -G^q_{\omega, \ell{+}\frac12}(r) F^p_{\omega, \ell{+}\frac12}(r')           \left[(\ell{+}1){\bf 1} + \vec \Sigma \cdot \vec L\right]
+ F^q_{\omega, \ell{-}\frac12}(r) G^p_{\omega, \ell{-}\frac12}(r') \left[\ell{\bf 1} - \vec \Sigma \cdot \vec L\right] \; . 
\end{eqnarray}
Note that the subscripts $ \ell \pm \frc12 $ on the radial mode functions denote the value of $ j $ in the mode equations (\ref{eq:Feq}) and (\ref{eq:Geq}).
The operators $\vec L$ are defined in Eq.\ (\ref{eq:ang_op}), and always act on the unprimed variables $\theta$ and $\phi$.  

\section{WKB Approximation} 
\label{WKB}

The method presented in Sections \ref{Tmn_unren} and \ref{Tmnren} of computing the stress-energy tensor for the massless spin $ \frc12 $ field makes use of a WKB approximation for the solutions to the mode equations (\ref{eq:Feq}) and (\ref{eq:Geq}).  The WKB approximation can be obtained by writing 
\begin{mathletters}
\begin{eqnarray}
\label{eq:G_q}
	G^q_{\omega,j}(r) & = & \frac{1}{\sqrt{2 W(r)}} \exp{\left[ - \int W(r) \left( h/f \right)^{1/2} dr \right]} \\
\label{eq:G_p} 
	G^p_{\omega,j}(r) & = & \frac{1}{\sqrt{2 W(r)}} \exp{\left[ \int W(r) \left( h/f \right)^{1/2} dr \right]} \; .
\end{eqnarray} 
\end{mathletters}Substitution of Eqs.\ (\ref{eq:G_q}) and (\ref{eq:G_p}) into Eq. (\ref{eq:Feq}) gives
\begin{mathletters}
\begin{eqnarray}
\label{eq:F_q}
	F^q_{\omega,j}(r) & = & \frac{G^q_{\omega,j}(r)}{\omega} \left( -W(r) - \frac{f^{1/2}}{h^{1/2}} \frac{W'(r)}{2 W(r)} + (j{+}\frc12) \frac{f^{1/2}}{r} \right) \\ 
\label{eq:F_p}
	F^p_{\omega,j}(r) & = & \frac{G^p_{\omega,j}(r)}{\omega} \left( W(r) - \frac{f^{1/2}}{h^{1/2}} \frac{W'(r)}{2 W(r)} + (j{+}\frc12) \frac{f^{1/2}}{r} \right)
\end{eqnarray}
\end{mathletters}Note that Eqs.~(\ref{eq:G_q}--b) and (\ref{eq:F_q}--b) are normalized in this form so that the Wronskian condition (\ref{eq:wronskian}) is satisfied.
Then substituting Eqs.\ (\ref{eq:G_q}) and (\ref{eq:F_q}) into Eq.\ (\ref{eq:second_G}) or substituting Eqs.\ (\ref{eq:G_p}) and (\ref{eq:F_p}) into Eq.\ (\ref{eq:second_G}) gives the following equation for $ W(r) $:
\begin{eqnarray}
\label{eq:wkb}
	W^2(r) = \Omega^2(r) + \frac{f W(r)''}{2 h W(r)} -\frac{3 f W(r)'^2}{4 h W(r)^2} - \frac{f h' W(r)'}{4 h^2 W(r)} + \frac{f' W(r)'}{4 h W(r)}
\end{eqnarray}
with
\begin{eqnarray}
\label{eq:omega}
	\Omega^2(r) \equiv \omega^2 - \frac{(j{+}\frc12) f'}{2 h^{1/2} r} + \frac{(j{+}\frc12) f}{h^{1/2} r^2} + \frac{(j{+}\frc12)^2 f}{r^2} \; .
\end{eqnarray}
Eq.\ (\ref{eq:wkb}) can be solved iteratively.  The zeroth order solution is $ W(r)=\Omega(r) $.  The second order solution is
\begin{eqnarray}
	W(r)=\Omega + {1\over 4} \left[ \frac{f}{h \Omega^2} \Omega'' + \frac{1}{2 \Omega^2} \left( \frac{f'}{h} - \frac{f h'}{h^2} \right) \Omega'-\frac{3 f \Omega'^2}{2 h \Omega^3} \right] ,
\end{eqnarray}
and so forth.

\section{Point Splitting}
\label{rev_pt_split}
As is the case for most quantities in quantum field theory, the stress-energy tensor for quantized fields is divergent, and must be renormalized.  The method of renormalization which we choose is that of point splitting.  In this method~\cite{dewitt} one expands the expression for the stress-energy tensor in powers of the geodetic interval $ \sigma(x,x') $ and its derivatives $ \sigma^{; \nu}(x,x')$.    The quantity $ \sigma(x,x') $ is defined as one-half of the square of the proper distance along the shortest geodesic that joins the points $ x $ and $ x' $.  It has the property
\begin{eqnarray}
    \sigma(x,x') = \frc12 ( \sigma_\mu \sigma^\mu )
\end{eqnarray}
with
\begin{eqnarray}
\label{eq:sigma_mu}
	\sigma^\mu \equiv \sigma^{;\mu} \; .
\end{eqnarray}
One then subtracts the results from the unrenormalized stress-energy tensor and allows the points to come together.  Christensen \cite{christensen1,christensen2} has calculated the renormalization counterterms for the stress-energy tensor for free quantized fields of spin $0$, $ \frc12 $, and $1$.  
To actually implement the point splitting renormalization scheme, one usually chooses a set of coordinates, and chooses the way in which the points are split.  The geodetic interval $ \sigma $ and its derivatives are then expanded in powers of $ x^\mu-x^{\prime\mu} $.

As is shown in Section \ref{Tmn_unren}, the unrenormalized stress-energy tensor may be written in terms of covariant derivatives of the Euclidean Green's function.  It is made up of terms which are bi-spinors and sometimes also bi-vectors.  A bi-spinor is an object that transforms like a spinor at $ x $ and a spinor at $ x' $ under Lorentz transformations (or Euclidean rotations) and 
a bi-vector is an object that transforms like a vector at $ x $ and a vector at $ x' $.

In general, a spinor at $ x' $ is parallel transported to a spinor at $ x $ through the use of the bi-spinor of parallel transport $ I(x,x') $.  A vector at $ x' $ is parallel transported to a vector at $ x $ through use of the bi-vector of parallel transport $ {g^\mu}_{\lambda'} $.   One may show that an object which is both a vector and a spinor is parallel transported by $ {g^\mu}_{\lambda'} I(x,x') $.  For example,
\begin{eqnarray}
	\left(\gamma^\mu\psi(x)\right)_\parallel = {g^\mu}_{\lambda'} I(x,x') \gamma^{\lambda'} \psi(x')
\end{eqnarray}
is a spinor and vector at $ x $, and a scalar at $ x' $.
We show in Appendix \ref{Ixpx} that 
to parallel transport $ \overline{\psi}(x') $ to $ \overline{\psi}(x) $, one must multiply on the right hand side by $ I(x',x) $, where $ I(x,x') I(x',x) = {\bf{1}} $.  Then
\begin{eqnarray}
	\left(\overline{\psi}(x)\gamma^\mu\right)_{\parallel} = {g^\mu}_{\lambda'} \overline{\psi}(x') \gamma^{\lambda'} I(x',x)
\label{eq:psibar} 
\end{eqnarray}
is again a spinor and vector at $ x $ and a scalar at $ x' $.

We shall split the points along the time direction so that
\begin{eqnarray}
\label{eq:pt_splitting}
    \epsilon \equiv t-t' , \; \; r=r', \; \; \theta=\theta', \; \; \phi=\phi' \; .
\end{eqnarray}
The quantities $ \sigma^\mu $ can then be expanded in powers of $ \epsilon $.
The result for a general static spherically symmetric spacetime is \cite{ahs} 
\begin{eqnarray}
\label{eq:sigma_epsilon}
     \nonumber \sigma^t & = & \epsilon + \frac{f'^2}{24 f h} \epsilon^3 - \frac{1}{120} \left( \frac{f'^4}{8 f^2 h^2} + \frac{3}{16} \frac{f'^3 h'}{f h^3} -\frac{3}{8} \frac{f'^2 f''}{f h^2} \right) \epsilon^5 + O(\epsilon^7) \\[.1cm] \nonumber 
\sigma^r & = & -\frac{f'}{4 h} \epsilon^2 - \frac{1}{24} \left( -\frac{f'^2 h'}{8 h^3} + \frac{f' f''}{4 h^2} \right) \epsilon^4 + O(\epsilon^6) \\[.1cm] 
\sigma^{\theta} & = & \sigma^{\phi} = 0 \; .
\end{eqnarray}
The bi-vectors of parallel transport can also be expanded in powers of $ \epsilon $ with the result \cite{ahs}
\begin{eqnarray}
\label{eq:bivector_epsilon}
     \nonumber g^{tt'} & = & -\frac{1}{f} - \frac{f'^2}{8 f^2 h} \epsilon^2 + \left( \frac{f'^4}{384 f^3 h^2} - \frac{f'^2 f''}{96 f^2 h^2} + \frac{f'^3 h'}{192 f^2 h^3} \right) \epsilon^4 + O(\epsilon^6) \\[.1cm]
     \nonumber g^{tr'} & = & -g^{rt'} = -\frac{f'}{2 f h} \epsilon - \left( \frac{f'^3}{96 f^2 h^2} + \frac{f' f''}{48 f h^2} - \frac{f'^2 h'}{96 f h^3} \right) \epsilon^3 + O(\epsilon^5) \\[.1cm]
     \nonumber g^{rr'} & = & \frac{1}{h} + \frac{f'^2}{8 f h^2} \epsilon^2 - \left( \frac{f'^4}{384 f^2 h^3}- \frac{f'^2 f''}{96 f h^3} + \frac{f'^3 h'}{192 f h^4} \right) \epsilon^4 + O(\epsilon^6) \\[.1cm]
     \nonumber g^{\theta\theta'} &=& {1\over r^2}\\
     g^{\phi\phi'} &=&  {1\over r^2\sin^2\theta} \; .
\end{eqnarray}
All other components vanish.  In Appendix \ref{Ixpx}, we show that $ I(x',x) $ has the following expansion in terms of $ \epsilon $:
\begin{eqnarray}
    I(x',x) & = & I_1 {\bf 1} + \gamma^0 \tilde{\gamma} I_2 \; ,
\label{eq:Ixpx}
\end{eqnarray}
with
\begin{eqnarray}
\nonumber I_1 &=& 1 + \frac{f'^2}{16 f h} \frac{\epsilon^2}{2} + \left(-\frac{7}{256} \frac{f'^4}{f^2 h^2} - \frac{f'^3 h'}{32 f h^3} + \frac{f'^2 f''}{16 f h^2} \right)\frac{\epsilon^4}{24} +O(\epsilon^6)  \; , \\[.1cm]
    I_2 &=& \frac{f'}{4 f^{1/2} h^{1/2}} \, \epsilon +  \left(- \frac{1}{64} \frac{f'^3}{f^{3/2} h^{3/2}}- \frac{f'^2 h'}{32 f^{1/2} h^{5/2}} + \frac{f' f''}{16 f^{1/2} h^{3/2}} \right) \frac{\epsilon^3}{6} + O(\epsilon^5) \; .
\end{eqnarray} 

\section{An Unrenormalized Expression for $ \langle T_{\mu \nu} \rangle $ }
\label{Tmn_unren}
The stress-energy tensor for the classical spin $ \frc12 $ field is~\cite{birrell-davies}\footnote{Since our $ \gamma_{\mu} $ has the opposite sign as that in Ref. \cite{birrell-davies}, this introduces an apparent minus sign between our Eq.\ (\ref{eq:Tmn_minus_BD}) and Eq.\ (3.191) of Ref. \cite{birrell-davies}.} 
\begin{eqnarray}
\label{eq:Tmn_minus_BD}
	T_{\mu \nu}(x) = - \frac{i}{2} \left[ \overline{\psi}(x) \gamma_{( \mu} \nabla_{\nu )} \psi(x)-\left(\nabla_{( \mu} \overline{\psi} \left( x \right) \right) \gamma_{\nu )} \psi(x)\right]
\end{eqnarray}
where parentheses around tensor indices indicate symmetrization.  For example 
$ A_{(\mu \nu)} = \frc12 \left( A_{\mu \nu} + A_{\nu \mu} \right) $.
In Appendix \ref{TmnSE}, we show that, for the massless spin $ \frc12 $ field in a general static spherically symmetric spacetime,
\begin{eqnarray}
\label{eq:TmnSE}
\langle T_{\mu \nu}(x) \rangle_{\rm{unren}} = - \frc14 \lim_{x' \rightarrow x} {\rm Im} \left\{ {\rm Tr}
     \left[ \gamma_{(\mu} \left( \nabla_{\nu)} \left[ S_E + S_E^c \right] - {g_{\nu)}}^{\lambda'} \nabla_{\lambda'}
     \left[ S_E + S_E^c \right] \right) I(x',x) \right] \right\} \;,
\end{eqnarray}
where $S_E$ is understood to stand for $S_E(x,x')$.  In Appendix \ref{S1SE} the charge conjugation operation 
is defined for a matrix operator $U$ as
\begin{equation}
U^c \equiv C \left(\overline{U} \right)^T C^\dagger \;,
\end{equation}
where $^T$ means transpose and $C$ is the charge conjugation matrix, given by $C = i \gamma^2 \gamma^0$ in the Dirac representation.

One can write the components of $ \langle T_{\mu \nu} \rangle_{\rm unren}$ explicitly in terms of mode functions by substituting Eq.\ (\ref{eq:SE_1}) into Eq.\ (\ref{eq:TmnSE}).  In Appendix \ref{S1SE}, it is shown that when the
points are split in the time direction it is possible to write the results in terms of scalar functions multiplied by various combinations of Dirac matrices.\footnote{It is actually possible to do this for an arbitrary separation
of the points but the result is more complicated.}  This makes
it possible to compute the trace in Eq.\ (\ref{eq:TmnSE}) without explicitly solving the mode equations
or computing the mode sums.  The nonzero components of the unrenormalized stress-energy tensor are
\begin{eqnarray}
\label{eq:Tmn_unren}
\nonumber \langle {T_t}^t \rangle_{\rm unren} &=& 2 {\rm{Re}} \Biggl\{ \int {d\mu \over r^2} \Biggl[
     - \omega^2 \cos[ \omega (\tau{-}\tau') ] I_1 \left( g^{tt} {+} g^{tt'} \right) A_1
     - i \omega \sin[ \omega (\tau{-}\tau') ] I_1 g^{tr'} {h^{1/2} \over r} A_3 \\
\nonumber && - \, \omega^2 \cos[ \omega (\tau{-}\tau') ] I_2 g^{tr'} {h^{1/2} \over f^{1/2}} A_1
     - i \omega \sin[ \omega (\tau{-}\tau') ] I_2 \left( g^{tt}{-}g^{tt'} \right) {f' \over 4f^{1/2}h^{1/2}} A_1 \\
\nonumber && \left. + \omega \cos[ \omega (\tau{-}\tau') ] I_2 g^{tr'} {h^{1/2} \over r} A_2
     + i \omega \sin[ \omega (\tau{-}\tau') ] I_1 g^{tr'} \left({1 \over r} - {h^{1/2} \over r} + {f' \over 4f} \right)
     A_1 \Biggr] \right\} \\[.1cm]
\nonumber \langle {T_r}^r \rangle_{\rm unren} &=& 2 {\rm{Re}} \Biggl\{ \int {d \mu \; h^{1/2} \over r^2f^{1/2}} \Biggl[
     - i \omega \sin\left[ \omega (\tau{-}\tau') \right] I_2 \left( g^{rr}{-}g^{rr'} \right)
     \left( {1 \over r} - {h^{1/2} \over r} + {f' \over 4f} \right) A_1 \\
\nonumber && + \, i \omega \sin[ \omega (\tau{-}\tau') ] I_2 \left( g^{rr}{-}g^{rr'} \right){h^{1/2} \over r} A_3
     - \omega^2 \cos[ \omega (\tau{-}\tau') ] I_2 g^{rt'} A_1 \\
\nonumber && - \, \omega^2 \cos[ \omega (\tau{-}\tau') ] I_1 \left( g^{rr} {+} g^{rr'} \right) {h^{1/2} \over f^{1/2}} A_1
     + \omega \cos[ \omega (\tau{-}\tau') ] I_1 \left( g^{rr}{+}g^{rr'} \right) {h^{1/2} \over r} A_2 \\
\nonumber && + \, i \omega \sin{ \left[ \omega (\tau{-}\tau') \right]} I_1 g^{rt'} \frac{f'}{4 f^{1/2} h^{1/2}} A_1 \Biggr] 
     \Biggr\} \\[.1cm]
\nonumber \langle {T_\theta}^\theta \rangle_{\rm unren} &=& \langle {T_\phi}^\phi \rangle_{\rm unren} = 2 {\rm Re} \Biggl\{ 
     \int {d\mu \over r^3f^{1/2}} \Biggl[ -\omega \cos[ \omega (\tau{-}\tau') ] I_1 A_2
     - i \omega \sin[ \omega (\tau{-}\tau') ] I_2 A_3 \\
&& + \, i \omega \sin[ \omega (\tau{-}\tau') ] I_2 \left( {1 \over h^{1/2}} - 1 \right) A_1 \Biggr] \Biggr\}
\end{eqnarray}
where
\begin{eqnarray}
\label{eq:A_s}
\nonumber A_1 &=& \sum_{\ell=0}^\infty \left[ (\ell{+}1) F^q_{\omega,\ell+\frc12}(r) F^p_{\omega,\ell+\frc12}(r)
      - \ell \, G^q_{\omega ,\ell{-}\frc12}(r) G^p_{\omega ,\ell{-}\frc12}(r) + {r \over f^{1/2}} \right] \\
\nonumber A_2 &=& \sum_{\ell=0}^\infty \left[ \ell (\ell{+}1) \left( G^q_{\omega,\ell{+}\frc12}(r) 
     F^p_{\omega,\ell{+}\frc12}(r) + F^q_{\omega,\ell{-}\frc12}(r) G^p_{\omega ,\ell{-}\frc12}(r) \right)
     - {\ell (\ell{+}1) \over \omega} + {r^2 \omega \over 2f} \right] \\
\nonumber A_3 &=& \sum_{\ell=0}^\infty \left[ \ell (\ell{+}1) \left( F^q_{\omega,\ell{+}\frc12}(r)
     F^p_{\omega,\ell{+}\frc12}(r) + G^q_{\omega,\ell{-}\frc12}(r) G^p_{\omega ,\ell{-}\frc12}(r) \right)
     - {r \over 2f^{1/2}} + {r \over 2f^{1/2}h^{1/2}} \right. \\
&& \left. \hbox{\hskip2in} - {r^2 f' \over 4f^{3/2}h^{1/2}} \right] \;.
\end{eqnarray}
As in the scalar field case \cite{candelas,hc,howard,phi2,ahs}, there is a superficial divergence for large $\ell$, which arises because the points are split along the time direction.  As discussed in \cite{hc}, it cannot be physical since one knows that the Green's function is finite with the points split.  Since $ \delta(\tau-\tau') =0 $ if the points are split, one can add multiples of it and its derivatives to Eqs.~(\ref{eq:Tmn_unren}).  In particular one can add terms that will cancel the large $ \ell $ divergences in these equations.  In Eqs.\ (\ref{eq:A_s}), the terms that do not contain mode functions are these additional terms.

\section{Renormalized Expression for the Stress-Energy Tensor}
\label{Tmnren}
As mentioned previously, the renormalized stress-energy tensor is obtained by subtracting off the point splitting
counterterms and letting the points come together.  Thus  
\begin{eqnarray}
\label{eq:Tmnren_def}
	\langle T_{\mu \nu} \rangle_{\rm{ren}} = \lim_{x' \rightarrow x} \left( \langle T_{\mu \nu} \rangle_{\rm{unren}} -\langle T_{\mu \nu} \rangle_{\rm{DS}} \right) \;.
\end{eqnarray}
The DeWitt-Schwinger point splitting counterterms, $ \langle T_{\mu \nu} \rangle_{\rm{DS}} $ have been computed 
by Christensen~\cite{christensen2}.   In his derivation he eliminates odd powers of $ \sigma $ from the point splitting counterterms.  This results in the cancelation of terms with odd powers of $ \sigma^{\rho} $ in the numerator since
there are only even powers in the denominators of all terms.  However, the elimination of these terms results in the elimination of finite contributions to the stress-energy tensor, which must be included if it is to be conserved.

The linearly divergent terms can be derived through the use of equations (2.19)\ and (3.11)\ in Ref.\ \cite{christensen2}.  We find\footnote{Since we are working with a complex spin $ \frc12 $ field rather than the real field used in Ref.\ \cite{christensen2}, we effectively have twice as many fields.  Thus we have multiplied the counterterms in Ref.\ \cite{christensen2} by 2.} 
\begin{eqnarray}
\nonumber \langle T_{\mu \nu} \rangle_{\rm{DS,lin}} & = & \frac{1}{48 \pi^2} \left[ \left( - {R^{\alpha}}_{(\mu ; \, \nu)} +
  \frc12 g_{\mu \nu} R^{;\alpha} + R^{\alpha \hspace{.58cm} \lambda}_{\hspace{.2cm} (\mu \nu) \hspace{.18cm}; \, \lambda} \right) \frac{\sigma_{\alpha}}{\sigma^{\rho} \sigma_{\rho}}  
-\left( R^{;\alpha} -8 \frac{\sigma^{\beta}\sigma^{\gamma}}{\sigma^{\rho} \sigma_{\rho}} {R^{\alpha}}_{\beta; \gamma} \right) \frac{\sigma_{\alpha}\sigma_\mu\sigma_{\nu}}{(\sigma^{\rho} \sigma_{\rho})^2} \right. \\ \nonumber &-& \left.   \left( 4 {{R^{\alpha}}_{(\mu}}^{;\beta} -2 {R^{\alpha \beta}}_{; (\mu}+ 2 R^{\beta \hspace{.15cm} \alpha \hspace{.24cm} ;\lambda}_{\hspace{.17cm} \lambda \hspace{.15cm} (\mu} \right) \frac{\sigma_{\nu)} \sigma_{\alpha}\sigma_{\beta}}{(\sigma^{\rho} \sigma_{\rho})^2}
- 2 \left( R^{\alpha \beta; \gamma} g_{\mu \nu} + {{R^{\gamma}}_{(\mu \nu)}}^{\alpha; \beta} \right) \frac{\sigma_{\alpha}\sigma_{\beta}\sigma_{\gamma}}{(\sigma^{\rho} \sigma_{\rho})^2} \right] \; . \\
\label{eq:Tmn_lin}
\end{eqnarray}

If the mode equation could be solved exactly for a general static spherically symmetric spacetime, and the mode sums and integrals in Eqs.\ (\ref{eq:Tmn_unren}) and (\ref{eq:A_s}) could be computed analytically, it would be straightforward to compute $ \langle T_{\mu \nu} \rangle_{\rm{ren}} $.  However, for most static spherically symmetric spacetimes, this is not the case.  As a result, the limit in Eq.\ (\ref{eq:Tmnren_def})
is difficult to evaluate since, without a solution to the mode equations, one doesn't know how to expand $ \langle T_{\mu \nu} \rangle_{\rm unren} $ in powers of $x^\mu - x^{\prime\mu}$.  To get around this problem we use the WKB approximation
to isolate the divergences in $\langle T_{\mu\nu}\rangle_{\rm unren}$.  The procedure to do this is 
discussed in detail in Ref.\cite{ahs}.  We shall briefly review it here.

First one substitutes Eqs.\ (\ref{eq:G_q}), (\ref{eq:G_p}), (\ref{eq:F_q}), and (\ref{eq:F_p}) 
into Eqs.(\ref{eq:Tmn_unren}) and computes the fourth order WKB approximation for the
variable $W$.  This can be done by iterating Eq.\ (\ref{eq:wkb}) twice.  Then the sums over $\ell$ are computed
using the Plana sum formula~\cite{gv}
\begin{eqnarray}
\label{eq:plana}
	\sum_{j=k}^\infty g(j) = \frc12 g(k) +\int_k^{\infty} g(\tau) d\tau + i \int_0^{\infty} \frac{dt}{e^{2 \pi t}-1} \left[ g(k+ it) - g(k-it) \right]	
\end{eqnarray}
and the results are expanded in inverse powers of $\omega$. 
Only terms are retained that contribute terms of order $\omega^{-1}$ or less to the stress-energy tensor, since these are the ones that lead to ultraviolet
divergences when the points come together.  If the field is in a zero temperature state, a 
lower limit cutoff $\lambda$ is used in the integral over $\omega$ for
terms that are proportional to $\omega^{-1}$.  In this way the quantity we call 
$ \langle T_{\mu \nu} \rangle_{\rm{WKBdiv}} $ is computed.  We can add and subtract this quantity from the
right hand side of Eq.\ (\ref{eq:Tmnren_def}) with the result that
\begin{eqnarray}
\label{eq:Tmn_def}
\nonumber \langle T_{\mu \nu} \rangle_{\rm{ren}} & = & \left( \langle T_{\mu \nu} \rangle_{\rm{unren}} - \langle T_{\mu \nu} \rangle_{\rm{WKBdiv}} \right) + \left( \langle T_{\mu \nu} \rangle_{\rm{WKBdiv}}-\langle T_{\mu \nu} \rangle_{\rm{DS}} \right) \\
          & \equiv & \langle T_{\mu \nu} \rangle_{\rm{numeric}} + \langle T_{\mu \nu} \rangle_{\rm{analytic}} \; .
\end{eqnarray}
Because all of the divergences in $\langle T_{\mu \nu} \rangle_{\rm{unren}}$ and 
 $\langle T_{\mu \nu} \rangle_{\rm{DS}}$ are contained in $\langle T_{\mu \nu} \rangle_{\rm{WKBdiv}}$, both
$\langle T_{\mu \nu} \rangle_{\rm{numeric}}$ and $\langle T_{\mu \nu} \rangle_{\rm{analytic}}$ are finite.
The tensor
$\langle T_{\mu \nu} \rangle_{\rm{numeric}}$ must usually be computed numerically, while $\langle T_{\mu \nu} \rangle_{\rm{analytic}}$ may be computed analytically for an arbitrary static spherically symmetric spacetime.

To compute $\langle T_{\mu \nu} \rangle_{\rm{analytic}}$ one can use the formulas in Appendix~\ref{mode_sums}
to evaluate the integrals over $ \omega $ in $\langle T_{\mu \nu} \rangle_{\rm{WKBdiv}}$.  Then the analytic
continuation discussed in Appendix \ref{S1SE} is used to analytically continue to Lorentzian space.  
The details of this process are discussed in Appendix \ref{TmnSE}.  The resulting terms 
are proportional to various powers of $(t-t')$ or to $\ln(t-t')$.  The
point splitting counterterms are then also expanded in powers of $(t-t')$.  After the subtraction the limit
$t' \rightarrow t$ is computed.  As mentioned above,
for the zero temperature case, there is an infrared divergence in the $ \omega^{-1} $ term which can be handled with the introduction of an infrared cutoff, $ \lambda $.  
To compute $\langle T_{\mu \nu} \rangle_{\rm{numeric}}$ one can simply set $\tau' = \tau$ in the mode
integrals since this quantity is finite in the limit that the points come together.

Our results are:
\begin{eqnarray}
	\langle T_{\mu \nu} \rangle_{\rm{analytic}} = (T_{\mu \nu})_0 + (T_{\mu \nu})_{\rm{log}}
\end{eqnarray}
with
\begin{mathletters}
\begin{eqnarray}
\label{eq:Tmn_analytic}
\nonumber ({T_t}^t)_0&=& {1 \over 240\pi^2} \left[ - {7\kappa^4 \over 4f^2} + {5\kappa^2 \over 6r^2f} 
     - {5\kappa^2 \over 6r^2fh}  + {5\kappa^2f' \over 3rf^2h} + {5\kappa^2h' \over 6rfh^2} - {5\kappa^2f'h' \over 12f^2h^2}
     - {25\kappa^2{f'}^2 \over 24f^3h} + {5\kappa^2f'' \over 6f^2h} \right. \\
\nonumber && + \, {1 \over r^4} - {1 \over r^4h^2} + {43{f'}^2 \over 24r^2f^2h^2}
     - {5{f'}^2 \over 8r^2f^2h} - {19{f'}^3 \over 12rf^3h^2} + {77{f'}^4 \over 192f^4h^2} - {2h' \over r^3h^3}
     + {15f'h' \over 4r^2fh^3}\\
\nonumber && - \, {5f'h' \over 12r^2fh^2} -  {15{f'}^2h'\over 8rf^2h^3}
     + {{f'}^3h' \over 24f^3h^3} - {7{h'}^2 \over 4r^2h^4} - {3f'{h'}^2 \over 2rfh^4} - {19{f'}^2{h'}^2 \over 32f^2h^4}
     + {7{h'}^3 \over rh^5} - {7f'{h'}^3 \over 4fh^5} \\
\nonumber && - \, {5f'' \over 6r^2fh^2} + {5f'' \over 6r^2fh} + {5f'f'' \over 6rf^2h^2}
     - {{f'}^2f'' \over 12f^3h^2} - {2h'f'' \over rfh^3} + {9f'h'f'' \over 8f^2h^3} + {19{h'}^2f'' \over 8fh^4}
     - {3{f''}^2 \over 8f^2h^2} \\
\nonumber && + \, {h'' \over r^2h^3} + {f'h'' \over 2rfh^3} + {{f'}^2h'' \over 4f^2h^3} - {13h'h'' \over 2rh^4}
     + {13f'h'h'' \over 8fh^4} - {f''h'' \over fh^3} + {2f''' \over rfh^2} - {f'f''' \over 2f^2h^2} \\
&& \left. - \, {3h'f''' \over 2fh^3} + {h''' \over rh^3} - {f'h''' \over 4fh^3} + {f'''' \over 2fh^2} \right] \\ [.5cm]
\nonumber ({T_r}^r)_0 &=& {1 \over 240\pi^2} \left[ {7\kappa^4 \over 12f^2} - {5\kappa^2 \over 6r^2f}
     + {5\kappa^2 \over 6r^2fh} - {5\kappa^2f'\over 6rf^2h} + {5\kappa^2{f'}^2 \over 24f^3h} + {2f' \over r^3fh^2}
     - {5{f'}^2 \over 8r^2f^2h^2} + {5{f'}^2 \over 24r^2f^2h} \right. \\
\nonumber &&  - \, {7{f'}^3 \over 24rf^3h^2} + {{f'}^4 \over 192f^4h^2} + {f'h' \over 2r^2fh^3}
     - {{f'}^2h' \over 4rf^2h^3} + {{f'}^3h' \over 16f^3h^3} - {7f'{h'}^2 \over 4rfh^4} + {7{f'}^2{h'}^2 \over 32f^2h^4}
     - {2f'' \over r^2fh^2}\\
\nonumber && \left. + \, {2f'f'' \over rf^2h^2} - {{f'}^2f'' \over 8f^3h^2} + {h'f'' \over rfh^3}
     - {f'h'f'' \over 4f^2h^3} - {{f''}^2 \over 8f^2h^2} + {f'h'' \over rfh^3} - {{f'}^2h'' \over 8f^2h^3}
     - {f''' \over rfh^2} + {f'f''' \over 4f^2h^2} \right] \\  \\
\nonumber ({T_\theta}^\theta)_0 &=& ({T_\phi}^\phi)_0 = {1 \over 240\pi^2} \left[ {7\kappa^4 \over 12f^2}
     - {5\kappa^2f' \over 12rf^2h} + {5\kappa^2{f'}^2 \over 12f^3h} - {5\kappa^2h' \over 12rfh^2}
     + {5\kappa^2f'h' \over 24f^2h^2} - {5\kappa^2f'' \over 12f^2h} \right. \\
\nonumber && - \, {f' \over r^3fh^2} - {3{f'}^3 \over 16rf^3h^2} + {29{f'}^4 \over 64f^4h^2} - {3f'h' \over 2r^2fh^3}
     - {7{f'}^2h' \over 16rf^2h^3} + {73{f'}^3h' \over 96f^3h^3} - {3f'{h'}^2 \over 4rfh^4} \\
\nonumber && + \, {31{f'}^2{h'}^2 \over 32f^2h^4} + {7f'{h'}^3 \over 4fh^5}  + {f'' \over r^2fh^2}
     + {7f'f'' \over 12rf^2h^2} - {73{f'}^2f'' \over 48f^3h^2} + {11h'f'' \over 4rfh^3} - {2f'h'f'' \over f^2h^3} \\
\nonumber && - \, {19{h'}^2f'' \over 8fh^4} + {7{f''}^2 \over 8f^2h^2} +  {f'h'' \over 4rfh^3} - {3{f'}^2h'' \over 8f^2h^3}
      - {13f'h'h'' \over 8fh^4} + {f''h'' \over fh^3} - {3f''' \over 2rfh^2} + {3f'f''' \over 4f^2h^2}\\
&& \left. + \, {3h'f''' \over 2fh^3} + {f'h''' \over 4fh^3} - {f'''' \over 2fh^2} \right] \\ [.5cm]
({T_\mu}^\nu)_{\rm log} &=& -{1 \over 4\pi^2} {U_\mu}^\nu \; \ln \left(\mu f^{1/2} \over 2\lambda\right) \hskip3.45cm \hbox{if $T=0$,}\\
({T_\mu}^\nu)_{\rm log} &=& -{1 \over 4\pi^2} {U_\mu}^\nu \; \left[\ln\left(2\mu f^{1/2}\over\kappa\right) + C\right] \hskip2cm \hbox{if $T \ne 0$.}
\end{eqnarray}
\end{mathletters}In the final pair of expressions, $\mu$ is a renormalization parameter, $\lambda$ is 
the previously mentioned infrared cutoff (necessary only in the zero temperature case), $C$ is Euler's Constant, and
\begin{eqnarray}
\nonumber {U_\mu}^{\nu}  &=& \frc{1}{10} \left( {R_{\rho \mu \tau}}^{\nu} R^{\rho \tau} - \frc14 R^{\rho \tau} R_{\rho \tau} {g_{\mu}}^{\nu} \right) - \frc1{30} R \left( {R_{\mu}}^{\nu} - \frc14 R {g_{\mu}}^{\nu} \right) \\
&& \qquad + \frc1{20} ({{R_{\mu}}^{\nu})_{;\rho}}^{\rho} - \frc1{60} {R_{;\mu}}^{\nu} - \frc1{120} {R_{;\rho}}^{\rho} {g_{\mu}}^{\nu}  \; .
\end{eqnarray}
The zero temperature case is obtained by setting $ \kappa = 0 $ in the above equations.

Our results for $ \langle {T_{\mu}}^{\nu} \rangle_{\rm num} $ are: 
\begin{mathletters}
\begin{eqnarray}
\label{eq:Ttt_num}
\nonumber \langle {{T_t}^t} \rangle_{\rm num} &=& 4 \int {d\mu \over r^2f} \left[ \omega^2 A_1 - {r^2\omega^3 \over f}
     - \omega\, \left( {1 \over 12} - {1 \over 12h} + {rf' \over 6fh} - {5r^2{f'}^2 \over 48f^2h} +  {rh' \over 12h^2}
     - {r^2f'h' \over 24fh^2} \right. \right.\\
&& \left. \left. + \, {r^2f'' \over 12fh} \right) \right] + \int {d\mu \over \omega} {U_t}^t \\ [.3cm]
\label{eq:Trr_num}
\nonumber \langle {{T_r}^r} \rangle_{\rm num} &=& 4 \int {d\mu \over r^2f} \left[ {\omega A_2 f^{1/2}\over r}
     - \omega^2 A_1 + {r^2\omega^3 \over 3f} + \omega\, \left( {1 \over 12} - {1 \over 12h}
     + {rf' \over 12fh} - {r^2{f'}^2 \over 48f^2h} \right) \right] \\
&& + \, \int {d\mu \over \omega} {U_r}^r \\ [.3cm]
\label{eq:Tthth_num}
\nonumber \langle {T_\theta}^\theta \rangle_{\rm num} &=& \langle {T_\phi}^\phi \rangle_{\rm num} 
     = 2 \int {d\mu \over r^2f} \left[ -{\omega A_2 f^{1/2}\over r} + {2r^2\omega^3 \over 3f} + \omega\, \left( 
     {rf' \over 12fh} - {r^2{f'}^2 \over 12f^2h} + {rh' \over 12h^2} \right. \right. \\
& & \left. \left. - \, {r^2f'h' \over 24fh^2} + {r^2f'' \over 12fh} \right) \right]
     + \int {d\mu \over \omega} {U_\theta}^\theta \; ,
\end{eqnarray}
\end{mathletters}where it is understood that, in the zero temperature case, there is an infrared cutoff $\lambda$ in the final integral of each equation.  In these equations, the quantities $A_1$ and $A_2$ are given in Eqs.\ (\ref{eq:A_s}).

\section{Discussion}
\label{Discussion}

A method for computing the stress-energy tensor for massless spin $\frc12$ fields in static spherically symmetric spacetimes has been presented.  The full renormalized stress-energy tensor can be computed either
numerically or analytically in a particular spacetime geometry by first solving the radial mode
equations (\ref{eq:Feq}--b) using the correct boundary conditions for the problem at hand.  The result is then substituted  into Eqs.\ (\ref{eq:A_s}) and (\ref{eq:Ttt_num}--c), and the mode sums are computed.  
Then the analytic approximation is computed for the given spacetime 
geometry and added to the result.

For an analytic computation this is all that is necessary.  However for numerical computations there can be
a problem with getting the mode sums to converge to a sufficient accuracy since, from a 
practical point of view, one must truncate
both the mode sum in $\ell$ and the integral (or sum) over $\omega$ at large values of these quantities.
Fewer modes can be used if one makes use of the full WKB approximation for those modes with large
values of $\ell$ and/or $\omega$.  One way to do this would be simply to use it for modes beyond the
cutoffs.  Another is to use the full WKB approximation for the radial mode functions for all values of $\omega$ and $\ell$ in Eqs.~(\ref{eq:Ttt_num}--c).   
We shall call the resulting quantity
 $\langle T_{\mu\nu} \rangle_{\rm WKBfin}$.  It can be added and subtracted 
from the rest of the stress-energy tensor with the result that
\begin{eqnarray}
\langle T_{\mu\nu} \rangle_{\rm ren} &=& \langle T_{\mu\nu} \rangle_{\rm modes} + \langle T_{\mu\nu} \rangle_{\rm WKBfin}
                       + \langle T_{\mu\nu} \rangle_{\rm analytic} 
\nonumber \\
\langle T_{\mu\nu} \rangle_{\rm{modes}} &=& \langle T_{\mu\nu} \rangle_{\rm{unren}} - \langle T_{\mu\nu} \rangle_{\rm{WKB}} \nonumber \\
\langle T_{\mu\nu} \rangle_{\rm{WKBfin}} &=& \langle T_{\mu\nu} \rangle_{\rm{WKB}} - \langle T_{\mu\nu} \rangle_{\rm{WKBdiv}} \;.
\end{eqnarray}

The quantity $\langle T_{\mu\nu} \rangle_{\rm{WKBfin}}$ can sometimes be computed analytically but often must be computed numerically.  However, even if it must be computed numerically, one can go beyond the necessary numerical cutoff by expanding the WKB approximation for the modes with very large values of $\ell$ and $\omega$ in inverse powers of these quantities.  Then for each power, the series in $\ell$ can be summed or the integral (or series) in $\omega$ can be computed analytically.  Procedures for doing this are described in~\cite{ahs}. 

It is also possible to use $\langle T_{\mu\nu} \rangle_{\rm analytic}$ as an analytic approximation for the stress-energy
tensor.  Examination of Eq.\ (\ref{eq:Tmn_analytic}--e) shows that our result agrees with the analytic approximation
 derived by Frolov and Zel'nikov \cite{fz1}  when their expression is evaluated for a static spherically symmetric spacetime and their undetermined coefficients have the values $q_1^{(0)} = q_2^{(0)} = 0$ and $q_1^{(2)} = \frc1{144}$. Thus it also agrees with that of Brown, Ottewill, and Page~\cite{bop} in Schwarzschild spacetime.  

Work is in progress to generalize our method of computing $\langle T_{\mu\nu}\rangle$ to the case of 
a massive spin $ \frc12 $ field in a general static spherically symmetric spacetime.  We also plan to
use our method to numerically compute the stress-energy tensor for both massless and massive spin $ \frc12 $ 
fields in Schwarzschild and Reissner-Nordstr\"{o}m spacetimes.

\section*{Acknowledgments}
We would like to thank S.~M.~Christensen and W.~C.~Kerr for helpful discussions.
This work was supported in part by grants PHY-9800971 and
PHY-0070981 from the National Science Foundation.

\appendix 
\section{Comparison With the Usual Flat Space Green's Function}
\label{flat_green}
In this appendix we show that in flat space, our zero temperature Green's function $ S_E $ is equal to one derived by letting the Euclidean Dirac operator act on the usual scalar Green's function  $ {\cal{G}}_E $ 
multiplied by the unit matrix.  We do this by defining a new Green's function $G$ via the equation
\begin{equation}
\label{flat_green's}
\label{eq:SEGE}
     S_E \left(x,x' \right) = - \gamma^{\mu}_E \nabla_{\mu}^E G_E \left(x,x'\right)
\end{equation}
and showing that 
\begin{eqnarray}
\label{eq:GEGE}
	G_E(x,x') = {\cal{G}}_E(x,x') {\bf{1}} \; .
\end{eqnarray}

In flat space $G_E$ can be written in the form
\begin{eqnarray}
\label{eq:GE_us}
&&G_E(x,x') = - \int_{-\infty}^\infty {d \omega \, e^{i \omega \left( \tau{-}\tau' \right)} \over 2 \pi rr'}
      \sum_{j=\frc12}^\infty \sum_{m=-j}^j \biggl[ \theta(r{-}r')
      \pmatrix{A^{q,p}_{\omega,j,m} & 0 \cr 0 & A^{q,p}_{\omega,j,m}} + \theta(r'{-}r) \left( p \leftrightarrow q \right) \biggr] 
\end{eqnarray}
with
\begin{eqnarray}
A^{q,p}_{\omega,j,m} =  F_{\omega,j}^q(r)F_{\omega,j}^p(r')\Psi_{j,+}^m(\theta,\phi)\Psi_{j,+}^{m\dagger}(\theta',\phi')
     - G_{\omega,j}^q(r)G_{\omega,j}^p(r')\Psi_{j,-}^m(\theta,\phi)\Psi_{j,-}^{m\dagger}(\theta',\phi') \; .
\end{eqnarray}
To verify that this gives the correct $S_E$ in flat space one can first
substitute Eq.\ (\ref{eq:GE_us}) into Eq.\ (\ref{eq:SEGE}) with the result that
\begin{eqnarray}
\nonumber - \gamma^{\mu}_E \nabla^E_{\mu} G_E(x,x') &=& \pmatrix{ - \partial_{\tau} &
     i \tilde\sigma \left(  \partial_r - {1 \over r} \vec \sigma \cdot \vec L \right) \cr
     - i \tilde{\sigma} \left( \partial_r - {1 \over r} \vec \sigma \cdot \vec L \right) &
     \partial_{\tau} } G_E(x,x') \\[.1cm]
\nonumber &=& - i \int_{-\infty}^\infty {d \omega \, e^{i \omega \left( \tau{-}\tau' \right)} \over 2\pi rr'}
     \sum_{j=1/2}^{\infty} \sum_{m=-j}^j \Biggl[ \theta(r{-}r') \times \\
\nonumber &\times& \pmatrix{- \omega A^{q,p}_{\omega,j,m} &
     \tilde\sigma \left( \partial_r  - {1 \over r} (\vec \sigma \cdot \vec L + {\bf 1}) \right)  A^{q,p}_{\omega,j,m} \vspace{.2cm} \cr 
     - \tilde\sigma \left( \partial_r -{1 \over r} (\vec \sigma \cdot \vec L + {\bf 1}) \right) 
     A^{q,p}_{\omega,j,m} & \omega  A^{q,p}_{\omega,j,m} } \\
&&\qquad + \theta(r'{-}r) \left( p \leftrightarrow q \right) \Biggr] \; .
\end{eqnarray}
Next Eqs.\ (\ref{eq:ang_eigen_eqn}) can be used to show that 
\begin{eqnarray}
\nonumber \tilde\sigma \left(  \partial_r  - {1\over r}(\vec \sigma \cdot \vec L + {\bf 1})\right) A^{q,p}_{\omega,j,m}
     &=& \left( \partial_r - \frac{(j{+}\frc12)}{r} \right) F_{\omega,j}^q(r) F_{\omega,j}^p(r')
         \Psi_{j,-}^m (\theta,\phi)\Psi_{j,+}^{m\dagger}(\theta',\phi') \\
     &-& \left( \partial_r + \frac{(j{+}\frc12)}{r} \right) G_{\omega,j}^q(r) G_{\omega,j}^p(r')
         \Psi_{j,+}^m(\theta,\phi)\Psi_{j,-}^{m\dagger}(\theta',\phi') \; .
\end{eqnarray}
Finally through the use of the first order equations (\ref{eq:Feq}) and (\ref{eq:Geq}) one obtains Eq.\ (\ref{eq:SE}).

The usual form of the Euclidean Green's function in spherical coordinates in flat space is \cite{candelas,phi2}
\begin{eqnarray}
\label{eq:GE_usual}
     \nonumber {\cal G}_E(x,x') = \int_{-\infty}^\infty {d \omega \, e^{i \omega (\tau{-}\tau')} \over 8\pi^2 rr'} \sum_{\ell=0}^\infty (2\ell+1) & & \left[ \theta(r{-}r')u_{\omega,\ell}^q(r) u_{\omega,\ell}^p(r') \right. \\
     & & \left. + \, \theta(r'{-}r)u_{\omega,\ell}^p(r) u_{\omega,\ell}^q(r') \right] P_\ell(\cos\gamma)
\end{eqnarray}
where $ \cos{\gamma} $ is defined in Eq.\ (\ref{eq:cos_gamma}).  The radial mode functions $u_{\omega,\ell}$ satisfy the second-order differential equations
\begin{equation}
\partial_r^2 u_{\omega,\ell} - \omega^2u_{\omega,\ell} - {\ell^2+\ell \over r^2}u_{\omega,\ell} = 0 \; .
\end{equation}
They also satisfy the Wronskian condition
\begin{equation}
\left( \partial_r u_{\omega,\ell}^p \right) \, u_{\omega,\ell}^q  - \left( \partial_r u_{\omega,\ell}^q \right) \, u_{\omega ,\ell}^p = 1 \; .
\end{equation}
Here $u_{\omega,\ell}^p$ is finite at $r=0$ and $u_{\omega,\ell}^q$ is finite at $ r = \infty $.

To show that $ G_E(x,x') = {\cal{G}}_E(x,x') {\bf{1}} $, first note that the mode functions $ u_{\omega, \ell} $ can be chosen to satisfy the recursion relations
\begin{eqnarray}
\omega u^p_{\omega, \ell} = \partial_r u^p_{\omega,\ell{-}1} - {\ell \over r}u^p_{\omega,\ell{-}1} \hskip2cm
-\omega u^q_{\omega, \ell} = \partial_r u^q_{\omega,\ell{-}1} - {\ell \over r}u^q_{\omega,\ell{-}1} \; .
\end{eqnarray}
One can then easily show that both the differential equation and the Wronskian will be satisfied for $u_{\omega,\ell}$ if they are satisfied for $u_{\omega,\ell-1}$.  If we choose
\begin{eqnarray}
\label{eq:G_u's}
&&G_{\omega,j}^p = u_{\omega,j{+}\frc12}^p \hskip2cm G_{\omega,j}^q = u_{\omega,j{+}\frc12}^q \cr
&&F_{\omega,j}^p = u_{\omega,j{-}\frc12}^p \hskip2cm F_{\omega,j}^q = - u_{\omega,j{-}\frc12}^q \; ,
\end{eqnarray}
then the $F$'s and $G$'s will also satisfy Eqs.\ (\ref{eq:Feq}) and (\ref{eq:Geq}), 
and the appropriate boundary conditions.

Next note that substitution of Eqs.\ (\ref{eq:G_u's}) and (\ref{eq:psij's}) into Eq.\ (\ref{eq:GE_us}) gives
\begin{eqnarray}
\label{eq:GE_us_2}
\nonumber &&G_E(x,x') = {\bf{1}} \otimes \int_{-\infty}^\infty {d \omega \, e^{i \omega (\tau{-}\tau')} \over 2\pi rr'} \Biggl[\theta(r{-}r') \Biggl\{\sum_{\ell=0}^\infty \quad \sum_{m=-\ell{-}\frc12}^{\ell{+}\frc12} \Biggr[ u_{\omega,\ell}^q(r) u_{\omega,\ell}^p(r') \times  \\ [.2cm] \nonumber
&& \times \pmatrix{
     {\ell+{1\over2}+m \over 2\ell+1} Y_\ell^{m-{1\over2}}(\theta,\phi)Y_\ell^{m-{1\over2}}(\theta',\phi')^\ast &  
     {\sqrt{(\ell+{1\over2})^2-m^2} \over 2\ell+1} Y_\ell^{m-{1\over2}}(\theta,\phi)Y_\ell^{m+{1\over2}}(\theta',\phi')^\ast \vspace{.2cm} \cr
     {\sqrt{(\ell+{1\over2})^2-m^2} \over 2\ell+1} Y_\ell^{m+{1\over2}}(\theta,\phi)Y_\ell^{m-{1\over2}}(\theta',\phi')^\ast &
     {\ell+{1\over2}-m \over 2\ell+1} Y_\ell^{m+{1\over2}}(\theta,\phi)Y_\ell^{m+{1\over2}}(\theta',\phi')^\ast } \cr
&&\hbox{\hskip2.8in} + \quad \sum_{\ell=1}^\infty \quad \sum_{m=-\ell+\frc12}^{\ell - \frc12}u_{\omega,\ell}^q(r) u_{\omega,\ell}^p(r') \times \\ [.2cm] \nonumber
&& \times \pmatrix{
     {\ell+{1\over2}-m \over 2\ell+1} Y_\ell^{m-{1\over2}}(\theta,\phi)Y_\ell^{m-{1\over2}}(\theta',\phi')^\ast &  
     -{\sqrt{(\ell+{1\over2})^2-m^2} \over 2\ell+1} Y_\ell^{m-{1\over2}}(\theta,\phi)Y_\ell^{m+{1\over2}}(\theta',\phi')^\ast \vspace{.2cm} \cr
     -{\sqrt{(\ell+{1\over2})^2-m^2} \over 2\ell+1} Y_\ell^{m+{1\over2}}(\theta,\phi)Y_\ell^{m-{1\over2}}(\theta',\phi')^\ast &
     {\ell+{1\over2}+m \over 2\ell+1} Y_\ell^{m+{1\over2}}(\theta,\phi)Y_\ell^{m+{1\over2}}(\theta',\phi')^\ast } \Biggr\} \\ [.2cm]
&&\hbox{\hskip2in} + \; \theta(r'{-}r) \Biggl\{ \, p \leftrightarrow q \, \Biggr\} \Biggr] \; .
\end{eqnarray}
The off-diagonal terms can be shown to cancel.  Shifting the sums over $ m $ by $\pm\frc12$ can then be shown to give
\begin{eqnarray}
     G_E(x,x') =&& {\bf{1}} \otimes \int_{-\infty}^\infty {d \omega \, e^{i \omega (\tau{-}\tau')} \over 2\pi rr'} \Biggl[\theta(r{-}r') \biggl\{ \sum_{\ell=0}^\infty \quad \sum_{m=-\ell}^\ell u_{\omega,\ell}^q(r) u_{\omega,\ell}^p(r') \\[.2cm]
     \nonumber &&\qquad \pmatrix{Y_\ell^m(\theta,\phi)Y_\ell^m(\theta',\phi')^\ast & 0 \cr 0 & Y_\ell^m(\theta,\phi)Y_\ell^m(\theta',\phi')^\ast } \biggr\} \; + \; \theta(r'{-}r) \biggl\{ \, p \leftrightarrow q \, \biggr\} \Biggr] \; .
\end{eqnarray}
Using the identity (\ref{eq:addition_theorum}) one can rewrite this expression as
\begin{eqnarray}
\label{eq:GE_us_3}
     \nonumber G_E(x,x') = {\bf{1}} \int_{-\infty}^\infty {d \omega \, e^{i \omega (\tau{-}\tau')} \over 8\pi^2 rr'} \sum_{\ell=0}^\infty (2\ell+1) & & \left[ \theta(r{-}r')u_{\omega,\ell}^q(r) u_{\omega,\ell}^p(r') \right. \\
     & & \left. + \,  \theta(r'{-}r)u_{\omega,\ell}^p(r) u_{\omega,\ell}^q(r') \right] P_\ell(\cos\gamma)\; .
\end{eqnarray}
Comparison of Eqs.\ (\ref{eq:GE_us_3}) and (\ref{eq:GE_usual}) shows that Eq.\ (\ref{eq:GEGE}) is correct. 

Since in flat space we can obtain $ S_E $ from Eq.\ (\ref{eq:SEGE}), it is clear that our expression for $ S_E $ is equal to the usual flat space Green's function obtained from the relationship
\begin{eqnarray}
\label{eq:SEGE2}
     S_E \left(x,x' \right) = - \gamma^{\mu}_E \nabla_{\mu}^E \left( {\cal{G}}_E \left(x,x' \right) {\bf{1}} \right) \; .
\end{eqnarray}


\section{Bi-spinor of Parallel Transport}
\label{Ixpx}
The bi-spinor of parallel transport $ I(x,x') $ is required to parallel transport a spinor at $ x' $ 
to a spinor at $ x $~\cite{dewitt,christensen2,christensen_thesis}
\begin{eqnarray}
\label{eq:psi_I}
     \psi(x)_\parallel = I(x,x') \psi(x') \; .
\end{eqnarray}
 From this definition it is clear that $I(x,x) = {\bf 1}$.  Furthermore, if you parallel transport $\psi(x')$ from $x'$ to $x$, and then transport it back along the same geodesic, it should return to its original value, so
\begin{eqnarray}
\label{eq:I_identity}
     I(x',x)I(x,x') = {\bf 1} \; .
\end{eqnarray}

The bi-spinor can also be used to parallel transport $\overline\psi(x)$, namely,
\begin{equation}
     \overline\psi(x)_\parallel = \overline\psi(x') \overline{I}(x,x') \; .
\end{equation}
Since $\overline\psi(x) \psi(x)$ is a scalar quantity, we expect it to be unchanged by parallel transport, so that
\begin{equation}
\overline\psi(x') \, \psi(x') = \overline\psi(x)_\parallel  \, \psi(x)_\parallel
     = \overline\psi(x') \overline{I}(x,x') \, I(x,x') \psi(x') \; .
\end{equation}
Comparison with Eq.~(\ref{eq:I_identity}) shows that
\begin{equation}
\label{eq:I_Ibar}
    \overline{I}(x,x') = I(x',x)
\end{equation}
and therefore
\begin{equation}
\label{eq:psibar_I}
     \overline\psi(x)_\parallel = \overline\psi(x') I(x',x) \; .
\end{equation}

For an object of arbitrary spin to be parallel transported along a geodesic, the inner product of its covariant derivative and the tangent vector of the geodesice must be zero.  The tangent vector is equal to $\sigma^\mu$, which is defined in Eq.\ (\ref{eq:sigma_mu}).  Thus for a spin $\frc12$ field
\begin{eqnarray}
	\sigma^\mu \nabla_\mu \psi(x)_\parallel =0 \; .
\end{eqnarray}
Using Eq.\ (\ref{eq:psi_I}), one finds that~\cite{dewitt,christensen_thesis}
\begin{eqnarray}
\label{eq:I_def}
	 && \sigma^{\mu} \nabla_{\mu} I(x,x') = 0 \; .
\end{eqnarray}

In this paper we are renormalizing by splitting the points in the time direction as shown explicitly in Eq.\ (\ref{eq:pt_splitting}).  Thus we seek an expansion for $ I(x,x') $ in powers of $ \epsilon \equiv t-t' $.  However, 
Eq.\ (\ref{eq:I_def}) contains derivatives of $ I(x,x') $, so we cannot simply assume an expansion in powers of $ \epsilon $ and substitute into Eq.\ (\ref{eq:I_def}).  Instead it is necessary to consider an expansion for $ I(x,x') $ that is valid for an arbitrary separation of the points.  Such an expansion is of the form
\begin{eqnarray}
\label{eq:I_pt_split_x}
	I(x,x') & = & {\bf{1}} + I_{\alpha}(x) \epsilon^{\alpha} + \frac{1}{2!} I_{\alpha \beta}(x) \epsilon^{\alpha} \epsilon^{\beta} + \frac{1}{3!} I_{\alpha \beta \gamma}(x) \epsilon^{\alpha} \epsilon^{\beta} \epsilon^{\gamma} + \frac{1}{4!} I_{\alpha \beta \gamma \delta} (x) \epsilon^{\alpha} \epsilon^{\beta} \epsilon^{\gamma} \epsilon^{\delta} + 
\cdots
\end{eqnarray}
with $ \epsilon^{\alpha} \equiv x^{\alpha} - x^{\prime\alpha} $, and each coefficient $ I_{\alpha \beta \dots} $ symmetric in its indices.  Substitution of Eq.\ (\ref{eq:I_pt_split_x}) into Eq.\ (\ref{eq:I_def}) and use of Eq.\ (\ref{eq:der_def}) yields
\begin{eqnarray}
\label{eq:I_condition}
     \sigma^{\mu} \left[ \left( \Gamma_{\mu} + I_{\mu} \right) + \left( I_{\alpha,\mu} + \Gamma_{\mu} I_{\alpha} + I_{\alpha \mu} \right) \epsilon^{\alpha} + \cdots \right]=0 \; .
\end{eqnarray}
One can next expand $ \sigma^{\mu} $ in powers of $ \epsilon^{\alpha} $.  The procedure for doing this is described in \cite{christensen_thesis} and also in Appendix B of \cite{ahs}, where a misprint of \cite{christensen_thesis} is corrected.\footnote{The equation at the top of page 109 in Ref.~\cite{christensen_thesis} contains a sign error in the second term.}  One finds
\begin{eqnarray}
\label{eq:sigma_pt_split}
	\sigma^{\mu}(x,x') & = & s^{\mu}_{\alpha}(r) \epsilon^{\alpha} + \frac{1}{2!} s^{\mu}_{\alpha \beta}(r) \epsilon^{\alpha} \epsilon^{\beta} + \frac{1}{3!} s^{\mu}_{\alpha \beta \gamma}(r) \epsilon^{\alpha} \epsilon^{\beta} \epsilon^{\gamma} + \frac{1}{4!} s^{\mu}_{\alpha \beta \gamma \delta}(r) \epsilon^{\alpha} \epsilon^{\beta} \epsilon^{\gamma} \epsilon^{\delta} \; .
\end{eqnarray}
with
\begin{eqnarray*}
	s_{\alpha}^\mu & = & {g_\alpha}^\mu \\
      s_{\alpha \beta}^\mu & = & - \Gamma^\mu_{\alpha \beta} \\
      s_{\alpha \beta \gamma}^\mu & = & \Gamma^{\rho}_{\alpha \beta} \Gamma^\mu_{\rho \gamma} + \Gamma^\mu_{\alpha \beta, \gamma} \\
      s_{\alpha \beta \gamma \delta}^\mu & = & \frc13 \left[ \Gamma^{\rho}_{\alpha \beta} {R^\mu}_{\gamma \delta \rho} + \Gamma^{\rho}_{\alpha \gamma} {R^\mu}_{\beta \delta \rho} + \Gamma^{\rho}_{\alpha \delta} {R^\mu}_{\beta \gamma \rho} + \Gamma^{\rho}_{\beta \gamma} {R^\mu}_{\alpha \delta \rho} + \Gamma^{\rho}_{\beta \delta} {R^\mu}_{\alpha \gamma \rho}  + \Gamma^{\rho}_{\gamma \delta} \left( {R^\mu}_{\beta \rho \alpha} + {R^\mu}_{\alpha \rho \beta} \right)\right. \\
      & & \left. - \Gamma^{\rho}_{\alpha \beta, \delta} \Gamma^\mu_{\rho \gamma} - \Gamma^{\rho}_{\alpha \beta} \Gamma^\mu_{\rho \gamma, \delta} - \Gamma^{\rho}_{\alpha \beta, \gamma} \Gamma^\mu_{\rho \delta} - \Gamma^{\rho}_{\alpha \beta} \Gamma^{\tau}_{\rho \gamma} \Gamma^\mu_{\tau \delta}  - \Gamma^\mu_{\alpha \beta,\gamma \delta} \right] \;.
\end{eqnarray*}
Substituting Eqs.\ (\ref{eq:I_pt_split_x}) and (\ref{eq:sigma_pt_split}) into Eq.\ (\ref{eq:I_condition}) results
in an equation that contains terms proportional to $ \epsilon^{\alpha} $ , $ \epsilon^\alpha \epsilon^\beta$, $\ldots \;.$ 
Setting the coefficients of these terms to zero results
in a series of equations that can be solved for $I_\alpha$, $I_{\alpha \beta}$, and so forth.
Since we ultimately want the points to be split in the time direction we only need the quantities
$I_t$, $I_{tt}$, $I_{ttt}$, and $I_{tttt}$ in Eq.\ (\ref{eq:I_pt_split_x}).  However to obtain expressions for these it is also necessary to additionally compute $I_r$, $ I_{tr} $, $I_{rr}$, $ I_{ttr} $, $ I_{trr}$, and $I_{rrr}$.  We find 
for these latter terms that
\begin{eqnarray}
\label{eq:a_few_Irs}
\nonumber I_r &=& I_{rr} = I_{rrr} = 0 \\
\nonumber I_{tr} &=& {\gamma^0 \tilde\gamma \over f^{1/2} h^{1/2}} \left( - {{f'}^2 \over 16f} - {f'h' \over 16h}
     + {f'' \over 8} \right) \\
\nonumber I_{ttr} &=& {{f'}^3 \over 32f^2h} + {{f'}^2h' \over 32fh^2} - {f'f'' \over 16fh} \\
I_{trr} &=& {\gamma^0 \tilde\gamma \over f^{1/2} h^{1/2}} \left( - {f'^3 \over 12 f^2} - {11f'^2h' \over 192fh}
     - {11f'h'^2 \over 192h^2} + {f'f'' \over 6f} + {7 h'f'' \over 96h} + {f'h'' \over 24h} - {f''' \over 12} \right) \; .
\end{eqnarray}
The terms that survive in Eq.\ (\ref{eq:I_pt_split_x}) when the points are split in the time direction are
\begin{eqnarray}
\nonumber I_t &=&-\frac{f' \gamma^0 \tilde{\gamma}}{4 f^{1/2} h^{1/2}} \\
\nonumber I_{tt} & =& \frac{f'^2}{16 f h} \\
\nonumber I_{ttt}&=& {\gamma^0 \tilde\gamma \over f^{1/2} h^{1/2}} \left({f'^3 \over 64fh} + {f'^2h' \over 32h^2}
     - {f'f'' \over 16h}\right) \\
I_{tttt} & = & -\frac{7}{256} \frac{f'^4}{f^2 h^2} - \frac{f'^3 h'}{32 f h^3} + \frac{f'^2 f''}{16 f h^2} \; .
\end{eqnarray}
Thus the bi-spinor of parallel transport, in the time direction, is given by
\begin{eqnarray}
\label{eq:expansion_Ixxp}
I(x,x') & = & {\bf{1}} + I_t \epsilon + \frc12 I_{tt} {\epsilon}^2 + \frc16 I_{ttt} {\epsilon}^3
     + \frc1{24} I_{tttt} {\epsilon}^4 + O(\epsilon^5) \; .
\end{eqnarray}
Similarly,
\begin{eqnarray}
\label{eq:expansion_Ixpx}
I(x',x) & = & {\bf{1}} - I_t \epsilon + \frc12 I_{tt} {\epsilon}^2 - \frc16 I_{ttt} {\epsilon}^3
     + \frc1{24} I_{tttt} {\epsilon}^4 + O(\epsilon^5) \; .
\end{eqnarray}

\section{Derivation of An Expression for the Unrenormalized Stress-Energy Tensor in Terms of the Hadamard Green's Function}
\label{Tmn_S1}

In this appendix, we use a method similar to that followed by Christensen \cite{christensen2}, 
to derive an expression for $ \langle T_{\mu \nu} \rangle_{\rm{unren}} $ in terms of the spinor Hadamard Green's function  $ S^{(1)}(x,x') $.  However, we work with a complex field instead of the real one treated by Christensen, and we use different conventions for the Dirac matrices.  

The action for a massless, complex spin $ \frc12 $ field can be written \cite{birrell-davies}
\begin{eqnarray}
	S= \int d^4x \sqrt{-g} \; \frac{i}{2} \left[ \overline{\psi}(x) \; \gamma^\mu \nabla_\mu \psi(x) - \left(\nabla_\mu \overline{\psi}(x)\right) \; \gamma^\mu \psi(x) \right] \; .
\end{eqnarray}
The stress-energy tensor for the classical spin $ \frc12 $ field is obtained as the variation of this action with respect to the metric
\begin{equation} 
T_{\mu \nu}(x) = - \frac{2}{\sqrt{-g(x)}} \frac{\delta S}{\delta g^{\mu \nu}(x)} = - \frac{i}{2} \left\{ \overline{\psi}(x) \gamma_{( \mu} \nabla_{\nu )} \psi(x)-(\nabla_{( \mu} \overline{\psi}(x)) \gamma_{\nu )} \psi(x)\right\} \; .
\end{equation}
Using the anticommutivity of spin $ \frc12 $ fields and the cyclic property of the trace, one can show that 
\begin{eqnarray}
T_{\mu \nu}(x) = \frac{i}{4} {\rm{Tr}} \left( \gamma_{(\mu} \left[ \nabla_{\nu)} \psi(x), \overline{\psi}(x)\right] + \gamma_{(\mu} \left[ \nabla_{\nu)} \overline{\psi}(x), \psi(x) \right] \right) \; .
\end{eqnarray}
Next the points can be split symmetrically, so that
\begin{eqnarray}
\label{eq:Tmn_psi}
\nonumber T_{\mu \nu}(x) &=& \frc{i}{8}  \lim_{x' \rightarrow x} {\rm{Tr}} \left\{
     \gamma_{(\mu} \left( \nabla_{\nu)} \left[ \psi(x), \overline{\psi}(x') \right] \right) I(x',x)
   + \gamma_{(\mu} I(x,x') \nabla_{\nu)} \left[ \overline{\psi}(x), \psi(x') \right] \right. \\
& &  \left. + \, \gamma_{(\mu} {g_{\nu)}}^{\lambda'} \left( \nabla_{\lambda'}
     \left[ \overline{\psi}(x'), \psi(x) \right] \right) I(x',x)
   + \gamma_{(\mu} {g_{\nu)}}^{\lambda'} I(x,x') \nabla_{\lambda'} \left[ \psi(x'), \overline{\psi}(x) \right] \right\} \; .
\end{eqnarray}
Here we have used the bi-vector of parallel transport $ {g_\mu}^{\nu'} $ and the bi-spinors of parallel transport $ I(x,x') $ and $ I(x',x) $ to parallel transport the part of the expressions that are spinors and vectors at the point $ x' $ to 
the point $ x $ along the shortest geodecie connecting these points.

When the fermion field $ \psi $ is quantized we reinterpret (\ref{eq:Tmn_psi}) as an equation for the stress-energy operator.  One can then compute an expectation value of $ T_{\mu \nu}(x) $ in Eq.\ (\ref{eq:Tmn_psi}) with respect to whatever state the fields are in.  It is possible to write $ \langle T_{\mu \nu}(x) \rangle $ in terms of the Hadamard Green's function $ S^{(1)} $, which is defined as
\begin{eqnarray}
\label{eq:S1_def}
   S^{(1)}(x,x') \equiv \left\langle \left[ \psi(x), \overline{\psi}(x') \right] \right\rangle \; .
\end{eqnarray}
We find
\begin{eqnarray}
\label{eq:Tmn_S1}
\nonumber \left\langle T_{\mu \nu} \right\rangle &=& \frc{i}{8}  \lim_{x' \rightarrow x} {\rm{Tr}} \left\{
     \gamma_{(\mu} \left( \nabla_{\nu)} S^{(1)}(x,x') \right) I(x',x) 
   - \gamma_{(\mu} I(x,x') \nabla_{\nu)} S^{(1)}(x',x)  \right. \\
& & \qquad \left. - \, \gamma_{(\mu} {g_{\nu)}}^{\lambda'} \left( \nabla_{\lambda'} S^{(1)}(x,x') \right) I(x',x) 
   + \gamma_{(\mu} {g_{\nu)}}^{\lambda'} I(x,x') \nabla_{\lambda'} S^{(1)}(x',x) \right\} \; .
\end{eqnarray}

To simplify Eq.\ (\ref{eq:Tmn_S1}), first note that
\begin{eqnarray}
\label{eq:S1bar}
\overline{S}^{(1)}(x,x')= S^{(1)}(x',x) \; ,
\end{eqnarray}
which can easily be seen from Eq.\ (\ref{eq:S1_def}).  Since $S^{(1)}(x,x')$ is formed from the tensor product of $\psi(x)$ and $\overline{\psi}(x')$,
\begin{eqnarray}
\label{eq:S1_proof}
\overline{\nabla_{\nu} S^{(1)}(x,x')} & = & \overline{ \partial_{\nu} S^{(1)}(x,x') + \Gamma_\nu S^{(1)}(x,x')  }
     =  \partial_{\nu} S^{(1)}(x',x) + S^{(1)}(x',x) {\overline{\Gamma}}_\nu \; .
\end{eqnarray}
Using Eq.~(\ref{eq:Gamma_def}) and Eq.~(\ref{eq:sigma_ab_lor}), together with the identity $\overline\gamma_\mu = \gamma_\mu$, it is easy to show that $\overline\Gamma_\nu = - \Gamma_\nu$ so that
\begin{eqnarray}
\label{eq:S1_proof2}
\overline{\nabla_{\nu} S^{(1)}(x,x')}  =  \partial_{\nu} S^{(1)}(x',x) - S^{(1)}(x',x) \Gamma_\nu \; .
\end{eqnarray}
Since $S^{(1)}(x',x)$ is formed from the tensor product of $\psi(x')$ and $\overline{\psi}(x)$, and the covariant derivative of a barred spinor is
\begin{eqnarray}
\label{eq:der_bar_psi}
	\nabla_{\nu} \overline{\psi}(x) = \partial_{\nu} \overline{\psi}(x) - \overline{\psi}(x) \Gamma_{\nu} \;,
\end{eqnarray}
we identify the right hand of Eq. (\ref{eq:S1_proof2}) as simply a covariant derivative, so that
\begin{eqnarray}
\label{eq:S1_S1bar}
   \overline{ \nabla_{\nu} S^{(1)}(x,x')}&=& \nabla_\nu S^{(1)}(x',x) 
\end{eqnarray}
Barring this equation, and swapping $x \leftrightarrow x'$, we see that
\begin{eqnarray}
   \overline {\nabla_{\nu'} S^{(1)}(x,x')} = \nabla_{\nu'} S^{(1)}(x',x) 
\end{eqnarray}
Substituting these two equations into Eq.\ (\ref{eq:Tmn_S1}) we see that
\begin{eqnarray}
\label{eq:Tmn_S1bar}
\nonumber \langle T_{\mu \nu} \rangle = \frc{i}{8}  \lim_{x' \rightarrow x} & & {\rm{Tr}} \left\{
     \gamma_{(\mu} \left( \nabla_{\nu)} S^{(1)}(x,x') I(x',x) - I(x,x') \overline{\nabla_{\nu)} S^{(1)}(x,x')} \right. \right. \\
& & \left. \left. - \, {g_{\nu)}}^{\lambda'} \nabla_{\lambda'} S^{(1)}(x,x') I(x',x) + {g_{\nu)}}^{\lambda'} I(x,x')
  \overline{ \nabla_{\lambda'} S^{(1)}(x,x')} \right) \right\}
\end{eqnarray}
Using the cyclic property of the trace, together with Eq. (\ref{eq:I_Ibar}), this can be written
\begin{eqnarray}
\nonumber \langle T_{\mu \nu} \rangle &=& \frc{i}{8}  \lim_{x' \rightarrow x}  \left[ {\rm{Tr}} \left( \gamma_{(\mu} \nabla_{\nu)} S^{(1)}(x,x') I(x',x) \right) - {\rm{Tr}} \left( \overline{ \gamma_{(\mu} \nabla_{\nu)} S^{(1)}(x,x') I(x',x)} \right) \right. \\
& & \quad \left. - \, {\rm{Tr}} \left( \gamma_{(\mu} {g_{\nu)}}^{\lambda'} \nabla_{\lambda')} S^{(1)}(x,x') I(x',x) \right) + {\rm{Tr}} \left( \overline{ \gamma_{(\mu} {g_{\nu)}}^{\lambda'} \nabla_{\lambda')} S^{(1)}(x,x') I(x',x)} \right) \right]
\end{eqnarray}
One can then use the result for an arbitrary matrix $ A $ that $ {\rm{Tr}} \left[ \overline{A} \right]={\rm{Tr}} \left[A \right]^* $, to write
\begin{eqnarray}
\label{eq:TmnS1_2}
	\langle T_{\mu \nu} \rangle = - \frc{1}{4} \lim_{x' \rightarrow x} {\rm{Im}} \left\{ {\rm{Tr}} \left[ \gamma_{(\mu} \left( \nabla_{\nu)} S^{(1)}(x,x') - {g_{\nu)}}^{\lambda'} \nabla_{\lambda'} S^{(1)}(x,x') \right) I(x',x) \right] \right\}
\end{eqnarray}

\section{Relationship Between $ S^{(1)} $ and $ S_E $}
\label{S1SE}

In this section, we establish a relationship between the spinor Hadamard Green's function $ S^{(1)} $ and 
the Euclidean Green's function $S_E$.  This is done by first finding the relationship between $S^{(1)}$ and
the Feynman Green's function $ S_F $ and its charge conjugate $ S_F^c $.  Then a relationship is established 
between $S_F$ and $S_E$.


To establish a relationship between $ S^{(1)} $ and $ S_F $, one may start with the identity
\begin{eqnarray}
\label{eq:SF}
iS_F(x,x') = \frc12S^{(1)}(x,x')+ i \tilde S(x,x') \; ,
\end{eqnarray}
where
\begin{eqnarray}
\label{eq:Sbar_def}
i \tilde S(x,x') = \frc12\left[\theta(t{-}t')-\theta(t'{-}t)\right]\langle \{\psi(x), \overline \psi(x')\} \rangle \; .
\end{eqnarray}

It is possible to write $S^{(1)}$ in terms of $S_F$ by taking advantage of charge conjugation symmetry.  The charge conjugation operator acts on a solution $\psi(x)$ of the Dirac equation by replacing it with
\begin{eqnarray}
\psi^c(x) \equiv C (\overline{\psi}(x))^T = C \gamma^{0T} \psi^\ast(x) \; ,
\end{eqnarray}
where $^T$ means transpose and $C$ is the charge conjugation matrix which satisfies the constraints
\begin{eqnarray}
\label{eq:gamma_c_constraints}
C \gamma_a^T C^\ast = \gamma_a \qquad \hbox{and} \qquad CC^\ast = -{\bf 1} \; .
\end{eqnarray}
In the Dirac representation we can choose $ C = i \gamma^2\gamma^0 $.
It is then a relatively simple matter to show that, even in curved spacetime, if $\psi(x)$ is a solution of the Dirac equation (\ref{eq:dirac_equation}), then  $\psi^c$ is also a solution.  Furthermore, it is easy to see that $(\psi^c)^c = \psi$.

Suppose that in an arbitrary static spacetime all the positive frequency solutions are labeled $\psi_k(x)$.  Then the negative frequency solutions will be $\psi_k^c(x)$.  Hence when $\psi(x)$ is quantized, it can be written in terms of creation and annihilation operators as
\begin{eqnarray}
\label{eq:psi_raising}
	\psi(x) = \sum_k \left[ \psi_k(x) a_k + \psi_k^c(x) b_k^\dagger \right] \; .
\end{eqnarray}
Here the operators $a_k$ and $b_k$ satisfy the anticommutation relations
\begin{eqnarray}
\label{eq:anti_ak}
\{a_k,a_{k'}^\dagger\} = \{b_k,b_{k'}^\dagger\} = \delta_{kk'} \; ,
\end{eqnarray}
with all other anticommutators equal to zero.

If Eq.\ (\ref{eq:psi_raising}) is substitutde into Eq.\ (\ref{eq:S1_def}), then
\begin{equation}
\label{eq:S1_direct}
S^{(1)}(x,x') = \sum_k \sum_{k'} \left\langle\left[\psi_k(x)a_k + \psi_k^c(x) b_k^\dagger \; , \; \overline{\psi_{k'}}(x') a_{k'}^\dagger + \overline{\psi_{k'}^c}(x') b_{k'} \right] \right\rangle \; .
\end{equation}
For noninteracting fields in the vacuum state or a thermal state, the Hamiltonian and hence the density matrix is diagonal in the occupation number representation.  Therefore,
\begin{equation}
\label{eq:expect}
\langle a_{k'}^\dagger a_k \rangle = \delta_{kk'}f_k \; , \qquad \langle b_k^\dagger b_{k'} \rangle = \delta_{kk'} \bar f_k \; , \qquad \hbox{and} \qquad \langle a_k b_{k'} \rangle = \langle b_k^\dagger a_{k'}^\dagger \rangle = 0 \; ,
\end{equation}
where $f_k$ and $\bar f_k$ are the occupation functions for the particle and antiparticle states respectively.  Since the thermal state will, presumably, occupy the particle and antiparticle states equally, $f_k = \bar f_k$.  Then substituting Eq.\ (\ref{eq:anti_ak}) and Eq.\ (\ref{eq:expect}) in Eq.\ (\ref{eq:S1_direct}) gives
\begin{eqnarray}
\label{eq:S1_psi}
S^{(1)}(x,x')
&=&  \sum_k (1{-}2f_k) \left\{ \psi_k(x) \psi_{k}^\dagger(x')\gamma^0  \; - \; C\gamma^{0T}\psi_k^\ast(x) \psi_k^T(x')\gamma^{0\ast}C^\dagger \gamma^0 \right\} \; .
\end{eqnarray}
The charge conjugation operator acting on a matrix operator can be defined as
\begin{eqnarray}
S^c \equiv C (\overline{S})^T C^\dagger  = - C \gamma^{0T} S^\ast C^\dagger \gamma^0 \; .
\label{eq:charge_conj_op}
\end{eqnarray}
Substituting Eq.\ (\ref{eq:charge_conj_op}) into Eq.\ (\ref{eq:S1_psi}), and using Eq.\ (\ref{eq:gamma_c_constraints}) gives
\begin{eqnarray}
S^{(1)c}(x,x') 
&=&\sum_k (1{-}2f_k)\left\{-C\gamma^{0T} \psi_k^\ast(x) \psi_k^T(x') \gamma^{0\ast} C^\dagger \gamma^0 +\gamma^0 \gamma^{0\dagger} \psi_k(x) \psi_k^\dagger(x') \gamma^0\gamma^{0\dagger}\gamma^0 \right\} \; .
\end{eqnarray}
Then using $ (\gamma^{0})^{\dagger} = \gamma^0 $ and $ \gamma^0 \gamma^0 = {\bf{1}} $, one finds that $ S^{(1) \, c}(x,x') = S^{(1)}(x,x') $.
In a similar manner it can be shown that $\tilde{S}^c(x,x') = \tilde{S}(x,x') $.

Letting the charge conjugation operator act on Eq.\ (\ref{eq:SF}), and using the fact that both $S^{(1)}$ and $\tilde S$ are unchanged, gives
\begin{eqnarray}
-iS_F^c(x,x') = \frc12S^{(1)}(x,x') - i \tilde S(x,x') \; .
\end{eqnarray}
Combining this with Eq.\ (\ref{eq:SF}), gives
\begin{eqnarray}
\label{eq:S1SF}
	S^{(1)}(x,x') = iS_F(x,x')-iS_F^c(x,x') \; .
\end{eqnarray}


To establish a relation between $ S_E $ and $ S_F $ first note some results from flat space.  It is well known that the scalar Green's functions $ G_E $ and $ G_F $ are related by an analytic continuation \cite{birrell-davies}, so it is plausible to look for a relation of the same type between $ S_E $ and $ S_F $.  We take this analytic continuation to be
\begin{eqnarray}
\label{eq:analytic_continuation}
	\left( \tau{-}\tau' \right)^2 \rightarrow - \left( t {-} t' \right)^2 + i \varepsilon \; .
\end{eqnarray} 
To see that this analytic continuation is valid, consider the Euclidean Green's function $ G_E $ for the massless scalar field in flat space 
\begin{eqnarray}
\label{eq:GE_flat}
	G_E \left( \tau,\vec{x};\tau',\vec x \, ' \right) = \frac{1}{4 \pi^2} \left[ \left( \tau-\tau' \right)^2 + \left( \vec{x} - \vec x \, ' \right)^2 \right]^{-1} \; .
\end{eqnarray}
If it is analytically continued to the Lorentzian sector using Eq.\ (\ref{eq:analytic_continuation}), one finds that
\begin{eqnarray}
\label{eq:GF_flat}
	G_E \left( it,\vec{x};it',\vec x \, ' \right) = \frac{1}{4 \pi^2} {\cal{P}} \left( \frac{1}{- \left( t- t' \right)^2 + \left( \vec{x}{-}\vec x \, ' \right)^2} \right)
- \frac{i}{4 \pi} \delta \left[ - \left( t{-}t' \right)^2 + \left( \vec{x}{-}\vec x \, ' \right)^2 \right] \; ,
\end{eqnarray}
where ${\cal P}$ denotes the principal value and use has been made of the identity from complex analysis that $ ( \sigma + i \varepsilon )^{-1} = {\cal{P}} (\sigma^{-1}) - i\pi \delta ( \sigma ) $, with the limit $ \varepsilon \rightarrow 0^+ $ understood \cite{matthews-walker}.  From this equation it is easily seen that the correct relationship 
\begin{equation} 
G_E\left( it,\vec{x};it',\vec x \, ' \right) = i G_F\left( t,\vec{x};t',\vec x \, ' \right)  
\label{eq:GFGE}
\end{equation}
is obtained \cite{birrell-davies}.\footnote{Note that in Ref. \cite{birrell-davies} the relationship $ \tau = -it $ rather than $ \tau = i t $ is used.}

One may then obtain $ S_E $ from the flat space expression for $ G_E $ through the relation 
\begin{eqnarray}
     S_E \left( \tau,\vec{x}; \tau',\vec x \, ' \right) & = & - \gamma^{\mu}_E \nabla_{\mu}^E G_E \left( \tau,\vec{x}; \tau',\vec x \, ' \right) 
\end{eqnarray}
and $ S_F $ through the relation
\begin{eqnarray}
     S_F \left( t, \vec{x};t',\vec x \, ' \right) & = & i \gamma^{\mu} \nabla_{\mu} G_F \left( t , \vec{x};t' ,\vec x \, ' \right) \; .
\end{eqnarray}
Using Eqs.\ (\ref{eq:GFGE}), (\ref{eq:sigma_ab_euc}), (\ref{eq:gamma_euclidean}), and (\ref{eq:nabla_E_definition}) it is then easy to show that in flat space
\begin{eqnarray}
\label{eq:SESF}
	S_E \left( i t,\vec{x}; i t',\vec x \, ' \right) = i S_F\left( t , \vec{x}; t',\vec x \, ' \right)
\end{eqnarray}
where $ S_E\left( i t, \vec{x}; i t',\vec x \, ' \right) $ is shorthand for the analytic continuation of $ S_E\left( \tau,\vec{x}; \tau',\vec x \, ' \right) $ through the use of Eq.\ (\ref{eq:analytic_continuation}). 

For a general static spacetime Eq.\ (\ref{eq:SESF}) is consistent with the equations that $ S_E $ and $ S_F $ are known to satisfy.  To see this first note that the equation the Feynman Green's function $ S_F $ obeys is 
\begin{eqnarray}
\label{eq:SF_equation}
	i \gamma^{\mu} \nabla_{\mu} S_F\left( t , \vec{x}; t',\vec x \, ' \right) = \frac{\delta^4(x-x')}{\sqrt{-g}} \; ,
\end{eqnarray}
while $ S_E $ is a solution to Eq.\ (\ref{eq:SE_equation}).
If one substitutes $ \tau = i t $ into Eq.\ (\ref{eq:SE_equation}) and uses Eqs.\ (\ref{eq:gamma_euclidean}) and (\ref{eq:nabla_E_definition}), one finds\footnote{One can show, using a Wick rotation of the representation $ \delta ( \tau-\tau' ) = \int \frc{d\omega}{2 \pi} e^{i \omega (\tau-\tau')}$, that $ \delta \left( \tau - \tau' \right) = -i \delta \left( t- t' \right)$.}
\begin{eqnarray}
	i \gamma^{\mu} \nabla_{\mu} S_E\left( i t , \vec{x}; i t',\vec x \, ' \right) = i\, \frac{\delta^4(x-x')}{\sqrt{-g}} \; .
\end{eqnarray}
Thus $ S_E \left( i t , \vec{x}; i t',\vec x \, ' \right) $ satisfies the same equation as $ i S_F \left( t , \vec{x}; t',\vec x \, ' \right) $ does.  This provides strong evidence that the relationship (\ref{eq:SESF}) holds in a general static spacetime.

Finally we obtain the required relation between $S^{(1)}$ and $S_E$ by substituting Eq.~(\ref{eq:SESF})
into Eq.~(\ref{eq:S1SF}) to obtain
\begin{eqnarray}
\label{eq:S1SE}
	S^{(1)} \left( t, \vec{x}; t',\vec{x'} \right) = S_E \left( it,\vec{x};it',\vec x \, '\right) + S_E^c\left( it,\vec{x};it',\vec x \, '\right) \; .
\end{eqnarray}

\section{Stress-Energy Tensor Expressed in Terms of the Euclidean Green's Function}
\label{TmnSE}

In this appendix we derive explicit expressions for $ \langle T_{\mu \nu} \rangle_{\rm{unren}} $ and $ \langle T_{\mu \nu} \rangle_{\rm{numeric}} $.  We also show that both $ \langle T_{\mu \nu} \rangle_{\rm{numeric}} $ and $ \langle T_{\mu \nu} \rangle_{\rm{analytic}} $ are real quantities when the analytic continuation (\ref{eq:analytic_continuation}) is used.

The general expression for $\langle T_{\mu \nu} \rangle_{\rm{unren}}$ in terms of $S_E$ and its derivatives is obtained
by substituting Eq.~(\ref{eq:S1SE}) into Eq.~(\ref{eq:TmnS1_2}).  The result is
\begin{eqnarray}
\label{eq:Tmn_SE}
	\langle T_{\mu \nu} \rangle_{\rm{unren}} = - \frc{1}{4} \lim_{x' \rightarrow x} {\rm Im} \left\{ {\rm{Tr}} \left[ \gamma_{(\mu} \left( \nabla_{\nu)} \left[ S_E + S_E^c \right] - {g_{\nu)}}^{\lambda'} \nabla_{\lambda'} \left[ S_E + S_E^c \right] \right) I(x',x) \right] \right\} \;.
\end{eqnarray}

To derive an explicit expression for $ \langle T_{\mu \nu} \rangle_{\rm{unren}} $ in terms of sums and integrals over the mode functions $ F_{\omega,j} $ and $ G_{\omega,j} $, first substitute Eq.\ (\ref{eq:SE_1}) into Eq.\ (\ref{eq:Tmn_SE}) and compute the derivatives with the points split in an arbitrary direction.   Next set $ \vec{x} = \vec{x}\,'$
 but leave $ \tau $ and $ \tau' $ separated.  Then the various terms in the quantity $ \left( \nabla_{\nu)}  S_E - {g_{\nu)}}^{\lambda'} \nabla_{\lambda'}  S_E \right) $ in Eq.\ (\ref{eq:Tmn_SE}) can be written 
in terms of bi-scalars multiplied by Dirac matrices.  Specifically we find that 
\begin{eqnarray}
\label{eq:SE_space_together}
\nonumber \lim_{\vec x \, ' \rightarrow \vec{x}} S_E \left( \tau,\vec{x};\tau',\vec{x} \right) & = & - \gamma^0 D_1 \\
\nonumber \lim_{\vec x \, ' \rightarrow \vec{x}} \left[ {S_E\left( \tau,\vec{x};\tau',\vec x \, ' \right)}_{;t} \right] &=&
     - \gamma^0 D_4 + \frac{f'}{4 f^{1/2} h^{1/2}} \tilde{\gamma} D_1 \\
\nonumber \lim_{\vec x \, ' \rightarrow \vec{x}} \left[ {S_E\left( \tau,\vec{x};\tau',\vec x \, ' \right)}_{;t'} \right] &=& 
     \gamma^0 D_4 + \frac{f'}{4 f^{1/2} h^{1/2}} \tilde{\gamma} D_1 \\
\nonumber \lim_{\vec x \, ' \rightarrow \vec{x}} \left[ {S_E\left( \tau,\vec{x};\tau',\vec x \, ' \right)}_{;r} \right] &=& 
     - \gamma^0 {h^{1/2}\over r}D_3 + \tilde\gamma {h^{1/2} \over r} D_2 - \tilde\gamma {h^{1/2} \over f^{1/2}} D_4
     + \gamma^0 \left( {1 \over r} - {h^{1/2} \over r} + {f' \over 4f} \right) D_1 \\
\nonumber \lim_{\vec x \, ' \rightarrow \vec{x}} \left[ {S_E\left( \tau,\vec{x};\tau',\vec x \, ' \right)}_{;r'} \right] &=& 
     - \gamma^0 {h^{1/2}\over r}D_3 - \tilde\gamma {h^{1/2} \over r} D_2 + \tilde\gamma {h^{1/2} \over f^{1/2}} D_4
     + \gamma^0 \left( {1 \over r} - {h^{1/2} \over r} + {f' \over 4f} \right) D_1 \\
\nonumber \lim_{\vec x \, ' \rightarrow \vec{x}} \left[ {S_E\left( \tau,\vec{x};\tau',\vec x \, ' \right)}_{;\theta} \right] &=& 
     - \frac12 \left[ \gamma^0 D_3 +  \left( 1 - \frac{1}{h^{1/2}} \right) \gamma^0 D_1 + \tilde\gamma D_2 \right]
     \left( \gamma^1 \cos\phi + \gamma^2 \sin\phi \right) \gamma^3 \\
\lim_{\vec x \, ' \rightarrow \vec{x}} \left[ {S_E\left( \tau,\vec{x};\tau',\vec x \, ' \right)}_{; \theta'} \right] &=&
     \frac12 \left[ \gamma^0 D_3 + \left( 1 - \frac{1}{h^{1/2}} \right) \gamma^0 D_1 + \tilde\gamma D_2 \right] 
     \left( \gamma^1 \cos\phi + \gamma^2 \sin\phi \right) \gamma^3 
\end{eqnarray}
with
\begin{eqnarray}
\nonumber D_1 &=& \int {d\mu \over r^2 f^{1/2}} \omega \sin{\left[ \omega( \tau{-}\tau') \right] } A_1 \hspace{1.2cm}
          D_2 = \int {d\mu \over r^2 f^{1/2}} i \omega \cos{\left[ \omega( \tau{-}\tau') \right] } A_2 \\
          D_3 &=& \int {d\mu \over r^2 f^{1/2}} \omega \sin{\left[ \omega( \tau{-}\tau') \right] } A_3 \hspace{1.2cm} 
          D_4 = \int {d \mu \over r^2 f^{1/2}} i \omega^2 \cos{\left[ \omega( \tau{-}\tau') \right] } A_1 \; . 
\label{eq:D_s}
\end{eqnarray}
The quantities $A_1, A_2$, and $A_3$ are found in Eq.\ (\ref{eq:A_s}).
In deriving these equations use has been made of the mode equations (\ref{eq:Feq}) and (\ref{eq:Geq}).  In addition, some of the terms that occur in $S_E$ and its derivatives take the form
\begin{eqnarray}
\label{eq:bye_bye}
\sum_{\ell = 0}^\infty \left[(\ell{+}1) G^q_{\omega,\ell+\frc12}F^p_{\omega,\ell+\frc12} - \ell F^q_{\omega,\ell-\frc12}G^p_{\omega,\ell-\frc12} \right] \; .
\end{eqnarray}
These sums are superficially divergent, but in a manner similar to that discussed in Section \ref{Tmn_unren}, we can add terms proportional to $\delta(\tau - \tau')$ and its derivatives to $S_E$ and its derivatives in Eq.~(\ref{eq:SE_space_together}).  When the index on the first term is shifted, and the appropriate subtraction made, (\ref{eq:bye_bye}) takes the form
\begin{eqnarray}
\sum_{\ell = 0}^\infty \left[\ell\left( G^q_{\omega,\ell-\frc12}F^p_{\omega,\ell-\frc12} - F^q_{\omega,\ell-\frc12}G^p_{\omega,\ell-\frc12}\right) - {\ell \over \omega} \right] \; .
\end{eqnarray}
But by the Wronskian condition, Eq.~(\ref{eq:wronskian}), this vanishes.


To derive an equation for $ \langle T_{\mu\nu} \rangle_{\rm{ren}} $ one can substitute Eqs.\ (\ref{eq:SE_space_together})
into Eq.\ (\ref{eq:Tmn_SE}) and then compute the trace.  This is more easily accomplished if use is made of the
identity $ \left( \gamma^{\mu} \right)^c = - \gamma^{\mu} $, which holds for the Dirac representation
\cite{itzykson-zuber}.\footnote{This identity actually holds for any representation of the Dirac matrices for which $ C^{\dagger} = C^{-1}$.} The result is
\begin{eqnarray}
\label{eq:Tmn_u1}
\nonumber \langle {T_t}^t \rangle_{\rm unren} &=& f^{1/2} \, {\rm Im} \left\{ - \left( g^{tt}{+}g^{tt'} \right)
      \left( D_4{-}D_4^\ast \right) I_1
      - g^{tr'} \left({1 \over r} - {h^{1/2} \over r} + {f' \over 4f} \right) \left(D_1{-}D_1^\ast\right) I_1 \right. \\
\nonumber && + \, g^{tr'} {h^{1/2} \over r} \left( D_3{-}D_3^\ast \right) I_1
     + \left(g^{tt}{-}g^{tt'}\right) {f' \over 4f^{1/2} h^{1/2}} \left(D_1{-}D_1^\ast \right) I_2 \\
\nonumber & & \left. + \, g^{tr'} {h^{1/2} \over r} \left( D_2{-}D_2^\ast \right) I_2 
     - g^{tr'} {h^{1/2} \over f^{1/2}} \left( D_4{-}D_4^\ast \right) I_2 \right\} \\
\nonumber \langle {T_r}^r \rangle_{\rm unren} &=& {\rm Im} \left\{ \left(g^{rr}{-}g^{rr'}\right)
     \left({1 \over r} - {h^{1/2} \over r} + {f' \over 4f} \right) h^{1/2} \left(D_1{-}D_1^\ast \right) I_2
     - g^{rt'} h^{1/2} \left(D_4{-}D_4^\ast\right)I_2 \right. \\
\nonumber & &  - \, \left( g^{rr}{-}g^{rr'} \right) {h \over r} \left(D_3{-}D_3^\ast\right) I_2
     + \left( g^{rr}{+}g^{rr'} \right) {h \over r} \left(D_2{-}D_2^\ast \right) I_1 \\
\nonumber & & \left. - \, \left( g^{rr}{+}g^{rr'} \right) {h \over f^{1/2}} \left(D_4{-}D_4^\ast \right) I_1
     - g^{rt'} {f' \over 4 f^{1/2}} \left(D_1{-}D_1^\ast \right) I_1 \right\} \\
\langle {T_\theta}^\theta \rangle_{\rm unren} &=& {1 \over r} \, {\rm Im} \left\{ 
     \left(D_3{-}D_3^\ast\right) I_2 + \left(1{-}{1 \over h^{1/2}} \right) \left(D_1{-}D_1^\ast \right) I_2
     - \left( D_2{-}D_2^\ast \right) I_1 \right\} \; .
\end{eqnarray}
Since in each expression, we have terms of the form $(D_n{-}D_n^\ast)$, and we are taking the imaginary part, it is sufficient to replace these with $2D_n$.  This gives the expressions in Eq.~(\ref{eq:Tmn_unren}).

In principle one may solve the mode equations (\ref{eq:Feq}) and (\ref{eq:Geq}) exactly, compute the sums and 
integrals in Eqs.\ (\ref{eq:SE_space_together}), use the analytic continuation (\ref{eq:analytic_continuation}),
subtract off the point splitting counterterms displayed in Ref.~\cite{christensen2} and Eq.~(\ref{eq:Tmn_lin}), and take the limit $t' \rightarrow t$ to obtain a renormalized expression for the stress-energy tensor.  However, it is seldom
possible to solve the mode equations analytically and to analytically compute the sums and integrals in 
Eqs.\ (\ref{eq:SE_space_together}).  What can be done instead is to use the WKB approximation for the radial modes
to determine the divergence structure of $\langle T_{\mu\nu}\rangle_{\rm unren}$.  The procedure for doing this
has been discussed in Section VII.  The result is then substituted into Eq.\ (\ref{eq:Tmn_u1}) to
obtain the quantity $\langle T_{\mu\nu}\rangle_{\rm WKBdiv}$.  Then
\begin{eqnarray}
\nonumber	\langle T_{\mu \nu} \rangle_{\rm{unren}} & = & \left\{ \left( \langle T_{\mu \nu} \rangle_{\rm{unren}} - \langle T_{\mu \nu} \rangle_{\rm{WKBdiv}} \right) + \langle T_{\mu \nu} \rangle_{\rm{WKBdiv}} \right\} \\  & = & \lim_{t' \rightarrow t} \left\{ \langle T_{\mu \nu} \rangle_{\rm{num}}  + \langle T_{\mu \nu} \rangle_{\rm{WKBdiv}} \right\} \;.
\end{eqnarray}
Note that since $ \langle T_{\mu \nu} \rangle_{\rm{WKBdiv}} $ contains all of the divergences of $ \langle T_{\mu \nu} \rangle_{\rm{unren}} $, 
the quantity $ \langle T_{\mu \nu} \rangle_{\rm{num}} $ is explicitly finite in the limit $ \tau' \rightarrow \tau $.  As a result one may set $ \tau = \tau' $ before computing the mode sums and integrals in this quantity.  
Our explicit expressions for $ \langle T_{\mu \nu} \rangle_{\rm{num}} $ are given in Eqs.\ (\ref{eq:Ttt_num}--c).  

The quantity $ \langle T_{\mu \nu} \rangle_{\rm{WKBdiv}} $ can be computed analytically.  The resulting expression
contains terms of the form $ \left( \tau - \tau' \right)^{-n} $ with $ n=4, 3, 2, 1 , 0 $ as well as terms 
proportional to $ \ln{ \left( \tau - \tau' \right)} $. For even powers of $n$ the analytic continuation takes the
form
\begin{eqnarray}
\label{eq:even_case}
\frac{1}{\left( \tau - \tau' \right)^{n}} & = & \frac{1}{-\left( t - t' \right)^{n} + i \varepsilon}
     = -{\cal{P}} \left( \frac{1}{\left( t-t' \right)^n} \right) - i \pi \delta \left[ \left( t-t' \right)^n \right]  \; .
\end{eqnarray}
For odd powers of $n$ one may multiply and divide by $ \tau - \tau' $:
\begin{eqnarray}
	\frac{1}{\left( \tau - \tau' \right)^{n}} & = & \frac{\tau-\tau'}{\left( \tau - \tau' \right)^{n+1}} \; .
\end{eqnarray}
The numerator trivially contains no poles and therefore can be analytically continued using $ \tau = it $, while the remainder of the expression is treated in exactly the same manner as in Eq.\ (\ref{eq:even_case}). 
Upon subtracting the charge conjugate term, the terms in $ \langle T_{\mu \nu} \rangle_{\rm{WKBdiv}} $ that are proportional to delta functions cancel.  For the logarithmic terms
\begin{eqnarray}
\label{eq:log_part}
	\ln{ \left( \tau {-} \tau' \right)} = \frc12 \ln{ \left[ - \left( t{-}t' \right)^{2} + i \varepsilon \right]} = \frc12 \left[ \ln{ | \left( t{-}t' \right) ^{2} |} + i \pi \right]
\end{eqnarray}
so that again upon subtracting the charge conjugate, the imaginary part vanishes.  As a result, we find that $ \langle T_{\mu \nu} \rangle_{\rm{WKBdiv}} $ is real, and therefore that the difference $ \langle T_{\mu \nu} \rangle_{\rm{analytic}} = \lim_{t' \rightarrow t} \left( \langle T_{\mu \nu} \rangle_{\rm{WKBdiv}} - \langle T_{\mu \nu} \rangle_{\rm{DS}} \right) $ is both finite and real.  Our expressions for $\langle T_{\mu\nu}\rangle_{\rm analytic}$ are given in Eq.\ (\ref{eq:Tmn_analytic}--e).

\section{Computation of the Mode Sums in $ \langle T_{\mu\nu} \rangle_{\rm{WKBdiv}} $}
\label{mode_sums}
Here we compute the mode sums or integrals necessary to evaluate the quantity $\langle T_{\mu\nu} \rangle_{\rm WKBdiv}$ for non-zero and zero temperature states, respectively.
For the thermal case, Eq.\ (1.442.2) of Reference \cite{gr} gives:
\begin{equation}
\label{eq:first_sum}
	\kappa \sum_{n=1/2}^{\infty} \frac{\cos{ \left( n \kappa \epsilon_{\tau} \right)}}{n \kappa} = {1\over2}\ln\left[ \cot^2\left(\kappa \epsilon_{\tau} \over 4\right) \right] = -{1\over 2}\ln\left(\kappa^2\epsilon_\tau^2 \over 16\right) - {\kappa^2\epsilon_\tau^2 \over 48} - {7\kappa^4\epsilon_\tau^4 \over 23040} + O(\epsilon_\tau^6) \; ,
\end{equation}
where the sum is understood to be over half-integers $n=\frc12,\frc32,\ldots$.  For the sums and integrals we are interested in, $\epsilon_\tau = \tau - \tau'$.  Using Eq.~(\ref{eq:first_sum}) and taking derivatives with respect to $\epsilon_\tau$, we find that
\begin{eqnarray}
     \nonumber \kappa \sum_{n=1/2}^{\infty} \frac{\cos{ \left( n \kappa \epsilon_{\tau} \right)}}{n \kappa} &=& -{1\over2}\ln\left(\kappa^2\epsilon_\tau^2 \over 16\right) + O(\epsilon_\tau^2) \\
     \nonumber \kappa \sum_{n=1/2}^\infty \sin\left( n \kappa \epsilon_{\tau} \right) &=& {1 \over \epsilon_\tau} + O(\epsilon_\tau) \\
     \nonumber \kappa \sum_{n=1/2}^\infty n\kappa \, \cos\left( n \kappa \epsilon_{\tau} \right) &=& -{1 \over \epsilon_\tau^2} + {\kappa^2 \over 24} + O(\epsilon_\tau^2) \\
     \nonumber \kappa \sum_{n=1/2}^\infty (n\kappa)^2 \sin\left( n \kappa \epsilon_{\tau} \right) &=& -{2 \over \epsilon_\tau^3} + O(\epsilon_\tau) \\
      \kappa \sum_{n=1/2}^\infty (n\kappa)^3 \cos\left( n \kappa \epsilon_{\tau} \right) &=& {6 \over \epsilon_\tau^4} - {7 \kappa^4 \over 960} + O(\epsilon_\tau^2) \; .
\end{eqnarray}
In the zero temperature case the mode sums are the same as in \cite{ahs}.  It is necessary to introduce an infrared cutoff $ \lambda $ for the first integral.  Then
\begin{eqnarray}
	\nonumber \int_\lambda^\infty \frac{d \omega}{\omega} \cos(\omega\epsilon_\tau) & = & -\frac12 \ln\left(\lambda^2 \epsilon_\tau^2 \right) - C \\
      \nonumber \int_0^\infty d \omega \sin(\omega\epsilon_\tau) & = & {1 \over \epsilon_\tau} \\
      \nonumber \int_0^\infty \omega d \omega \cos(\omega \epsilon_\tau) & = & - {1 \over \epsilon_\tau^2} \\
      \nonumber \int_0^\infty \omega^2 d \omega \sin(\omega \epsilon_\tau) & = & - {2 \over \epsilon_\tau^3} \\
      \int_0^\infty \omega^3 d \omega \cos{ \left( \omega \epsilon_{\tau} \right)} & = & {6 \over \epsilon_\tau^4} \; ,
\end{eqnarray}
where $ C $ is Euler's constant.

\end{document}